# 2008-2009
# FRENCH COMPLEX SYSTEMS ROADMPAP

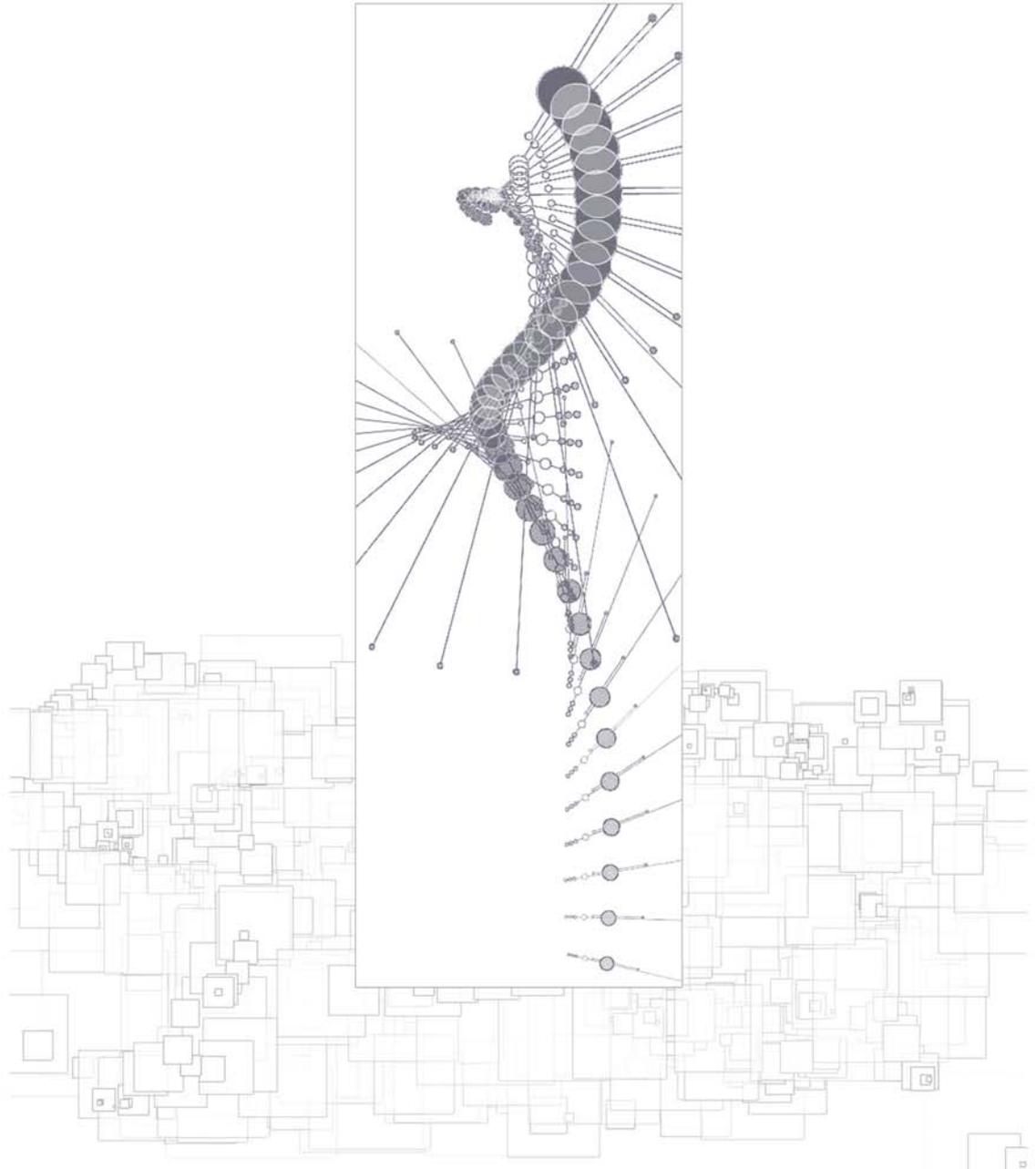
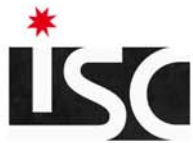
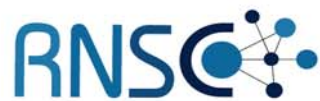
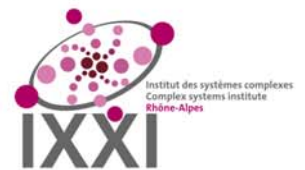

# TABLE OF CONTENT

# French Roadmap for Complex Systems 2008-2009 Edition

*This second issue of the French Complex Systems Roadmap is the outcome of the "Entretiens de Cargèse 2008", an interdisciplinary brainstorming session organized over one week in 2008, jointly by RNSC, ISC-PIF and IXXI. It capitalizes on the first roadmap and gathers contributions of more than 70 scientists from major French institutions.*

*The aim of this roadmap is to foster the coordination of the complex systems community on focused topics and questions, as well as to present contributions and challenges in the complex systems sciences and complexity science to the public, political and industrial spheres.*

## Editorial Committee

Paul Bourgine - Ecole Polytechnique

David Chavalarias - Institut des Systèmes Complexes de Paris Ile-de-France

Edith Perrier - Institut de Recherche pour le Développement

## Links

Réseau national des systèmes complexes (RNSC): http://rnsc.fr

Institut des Systèmes Complexes de Paris Île-de-France (ISC-PIF): http://iscpif.fr

Institut des Systèmes Complexes Rhône-Alpes (IXXI): http://ixxi.fr

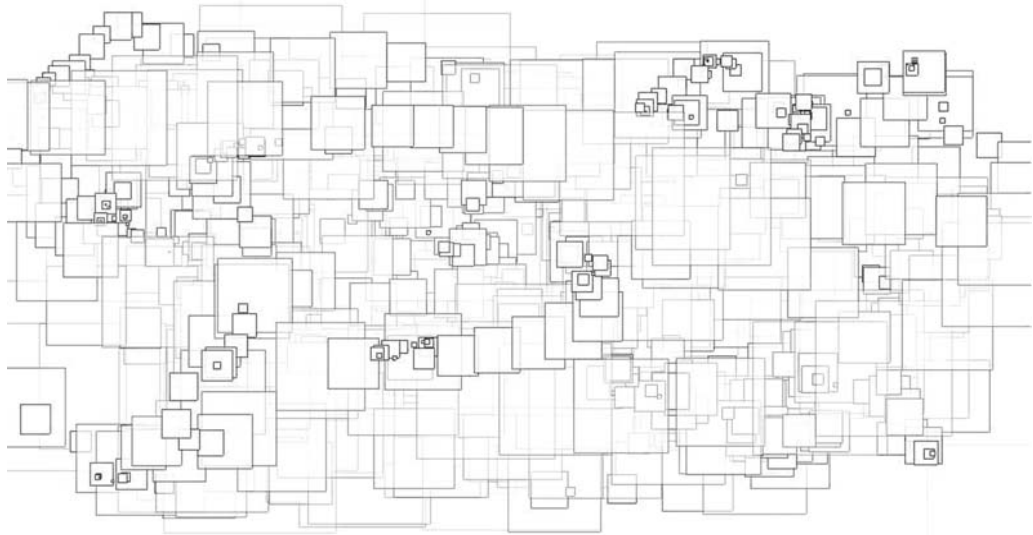

Illustrations for this roadmap have been generated with graphic environment provided Jared Tarbell (http://www.levitated.net)

# Contributors

**Note**: Some of the workgroups decided to designate a "reporter" for questions or topics represented by a large number of contributors. The role of reporters was to ensure that contributions were well balanced in their respective chapters, and that the final text was coherent.

Fréderic Amblard - Université de Toulouse

François Arlabosse - CYBEL

Pierre Auger - ISC-IRD-Geodes

Jean-Bernard Baillon - Université Paris 1

Olivier Barreteau - Cemagref

Pierre Baudot - ISC-PIF

Elisabeth Bouchaud - CEA

Soufian Ben Amor - Eurocontrol

Hugues Berry - INRIA

Cyrille Bertelle - LITIS UFR sciences technologies

Marc Berthod - INRIA

Guillaume Beslon - INSA Lyon, dept. Informatique

Giulio Biroli - CEA

Daniel Bonamy - CEA Saclay

Daniele Bourcier - CNRS

Paul Bourgine - Ecole Polytechnique

Nicolas Brodu - INRIA Rennes

Marc Bui - EPHE

Yves Burnod - INSERM

Bertrand Chapron - IFREMER

David Chavalarias - Institut des Systèmes Complexes de Paris Ile-de-France / CNRS

Catherine Christophe - INRA

Bruno Clément - Inserm

Jean-Louis Coatrieux - Université de Rennes 1

Jean-Philippe Cointet - Institut des Systèmes Complexes de Paris Ile-de-France / INRA

Valérie Dagrain - consultante technologie information

Katia Dauchot - CEA, Commissariat à l'energie atomique

Olivier Dauchot - CEA, Commissariat à l'energie atomique

François Daviaud - CEA

Silvia De Monte - ENS

Guillaume Deffuant - Meteo-France & U. Paris-Est Cemagref

Pierre Degond - CNRS

Jean-Paul Delahaye – LIFL - Université des Sciences et Technologies de Lille

René Doursat - Institut des Systèmes Complexes / CREA, Ecole Polytechnique

Francesco D'Ovidio - Ecole Normale Supérieure

Marc Dubois - CEA SPEC

Berengère Dubruelle - CEA

Marie Dutreix - Institut Curie

Robert Faivre - INRA

Emmanuel Farge - INSERM- University Paris 7 Denis Diderot

Patrick Flandrin - CNRS ENS Lyon

Sara Franceschelli - Paris Jussieu

Cédric Gaucherel - INRA

Jean-Pierre Gaudin - IEP

Michael Ghil - ENS Paris

Jean-Louis Giavitto - Université Evry

Francesco Ginelli - Institut des Systèmes Complexes de Paris Ile-de-France

Vincent Ginot - INRA

François Houllier - INRA

Bernard Hubert - INRA

Pablo Jensen - ENS Lyon

Ludovic Jullien - Paris VI University

Zoi Kapoula - CNRS

Daniel Krob - École Polytechnique

François Ladieu - CEA

Gabriel Lang - Agro Paris Tech

Chrsitophe Lavelle - IHES

André Le Bivic - Direction CNRS

Jean-Pierre Leca - Université Paris I-Panthéon-Sorbonne,

Christophe Lecerf - Ecole des Mines Ales

Pierre Legrain - CEA

Denis L'Hôte - CEA

Maud Loireau - IRD

Jean-Francois Mangin - CEA

Olivier Monga - IRD

Michel Morvan - Ecole normale supérieure de Lyon

Jean-Pierre Muller - CIRAD

Ioan Negrutiu - ENS Lyon

Edith Perrier - IRD

Nadine Peyreiras - Institut de Neurobiologie CNRS

Denise Pumain - Université Paris 1

Ovidiu Radulescu - Université de Rennes 1

Jean Sallantin - CNRS LIRMM

Eric Sanchis - Université Paul Sabatier

Daniel Schertzer - Météo France

Marc Schoenauer - INRIA

Michèle Sebag - CNRS

Eric Simonet - CNRS

Adrien Six - Université Pierre et Marie Curie - Paris 6

Fabien Tarissan - Institut des Systèmes Complexes de Paris Ile-de-France

Patrick Vincent

# Preamble

A "complex system" is in general any system comprised of a great number of heterogeneous entities, among which local interactions create multiple levels of collective structure and organization. Examples include natural systems ranging from bio-molecules and living cells to human social systems and the ecosphere, as well as sophisticated artificial systems such as the Internet, power grid or any large-scale distributed software system. A unique feature of complex systems, generally overlooked by traditional science, is the *emergence* of non-trivial superstructures which often dominate the system's behaviour and cannot be easily traced back to the properties of the constituent entities. Not only do higher emergent features of complex systems arise from lower-level interactions, but the global patterns that they create affect in turn these lower levels—a feedback loop sometimes called *immergence*. In many cases, complex systems possess striking properties of robustness against various large-scale and potentially disruptive perturbations. They have an inherent capacity to adapt and maintain their stability. Because complex systems require analysis at many different spatial and temporal scales, scientists face radically new challenges when trying to observe complex systems, in learning how to describe them effectively, and in developing original theories of their behaviour and control.

Complex systems demand an interdisciplinary approach, as the universal questions they raise find expression in le widely different systems across a broad spectrum of disciplines—from biology to computer networks to human societies. Moreover, the models and methods used to tackle these questions also belong to different disciplines—mainly computer science, mathematics and physics—and the standard methods of specialized domains rarely take into account the multiple-level viewpoint so needed in the context of complex systems, and attained only through a more integrated and interdisciplinary approach.

Two main kinds of interdisciplinary investigation can be envisioned. The first path involves working on a *topic* of research that is intrinsically multidisciplinary, for example "cognition," and posing various questions about the same topic from multiple and somewhat disconnected disciplinary viewpoints (from neuroscience, psychology, artificial intelligence, etc.). The second path consists in studying the same *question*, for example, "what causes synchronization?", in connection with different objects of research in different disciplines (statistical physics, chemistry, biology, electrical engineering, etc.). This second approach establishes the foundations of a true *science* of complex systems. However, the success of these two complementary approaches depends critically on the design of new protocols, new models and new formalisms for the reconstruction of emergent phenomena and dynamics at multiple scales. It is on the successful pursuit of this joint goal of (a) massive data acquisition on the basis of certain prior assumptions, and (b) reconstruction and modeling of such data, on which the future science of complex systems depends. There remains much to do in the theoretical domain in order to build concepts and models able to provide an elegant and meaningful explanation of the so-called "emergent" phenomena that characterize complex systems.

The aim of this roadmap is to identify a set of wide thematic domains for complex systems research over the next five years. Each domain is organized around a specific question or topic and proposes a relevant set of "grand challenges", i.e., clearly identifiable problems whose solution would stimulate significant progress in both theoretical methods and experimental strategies.

Theoretical questions issue immediately from the need to account for different levels of organization. In complex systems, individual behaviour leads to the emergence of collective organization and behaviour at higher levels, and these emergent structures in turn influence individual behaviour. This two-way causal influence raises important questions: what are the various levels of organization and what are their characteristic scales in space and time? How do reciprocal influences operate between the individual and collective behaviour? How can we simultaneously study multiple levels of organization, as is often required in problems in biology or social sciences? How can we efficiently characterize emergent structures? How can we understand the changing structures of emergent forms, their robustness or sensitivity to perturbations? Is it more important to study the attractors of a given dynamical system, or families of transient states? How can we understand slow and fast dynamics in an integrated way? What special emergent properties characterize those complex systems that are especially capable of adaptation in changing environments? During such adaptation, individual entities often appear and disappear, creating and destroying links in the system's graph of interactions. How can we understand the dynamics of these changing interactions and their relationship to the system's functions?

Other key questions arise out of the challenge of reconstructing system dynamics from data. These include questions related to the epistemic loop (the problem of moving from data to models and back to data, including model-driven data production), which is the source of very hard inverse problems. Other fundamental questions arise around the constitution of databases, or the selection and extraction of stylized facts from distributed and heterogeneous databases, or the deep problem of reconstructing appropriate dynamical models from incomplete, incorrect or redundant data.

Finally, some questions also emerge from efforts to govern or design complex systems, which one might think of as "complex systems engineering." On the basis of an incomplete reconstruction of dynamics based on data, how can we learn to steer a system's dynamics toward desirable consequences or at least keep itaway from regimes where it exhibits undesirable behaviour? How can one achieved a so-called "complex control," with controlling influences distributed on many distinct hierarchical levels in either a centralized or decentralized way? Finally, how is it possible to design complex artificial systems, using new techniques of multilevel control?

In addition to the questions just outlined, another key issue is the reaction of complex systems to perturbations, which can be weak for certain components or on certain scales of the system and strong for others. These effects, central to the prediction and control of complex systems and models, must be specifically studied. In addition, it is also important to develop both strategies for representing and extracting pertinent parameters and formalisms for modeling morphodynamics. Learning to successfully predict multiscale dynamics raises other important challenges, and will be required in going from controlled systems to governed systems in which control is less centralized and more distributed among hierarchical levels.

Grand challenges for complex systems research draw inspiration from a wide variety of complex phenomena arising in different scientific fields. Their presentation follows the hierarchy of organizational levels of complex systems, either natural, social or artificial. Understanding this hierarchy is itself a primary goal of complex systems science.

In modern physics, the understanding of collective behaviour and out-of-equilibrium fluctuations is increasingly important. Biology also faces complex behaviours at every level, in systems ranging from biological macromolecules and molecular systems through entire ecosystems. Indeed, the question of gaining an integrated understanding of the different scales of biological systems is probably one of the most difficult and exciting tasks for researchers in the next decade. Before we can hope to develop an integrated understanding of the full hierarchy of living systems, we must study and understand the integration between one level and the next. The hierarchy of levels includes cellular and sub-cellular spatiotemporal organization, and multicellular systems (integrating intracellular dynamics, such as gene regulation networks, with cell-cell signalling and bio-mechanical interactions), where the question of the impact of local perturbations on the stability and dynamics of multicellular organizations takes great importance. These systems underlie larger scale physiological functions, which emerge from sets of cells and tissues in complex interaction within a given environment. At the highest level, the understanding and control of ecosystems involves richly integrated interactions among living organisms in a given biotope.

In the context of the human and social sciences, too, the complex systems approach is rapidly assuming central importance (even if it is currently less developed than in biology). One crucial domain to be investigated is learning how the individual cognition of interacting agents leads to social cognition. Another important phenomenon, with particularly important societal consequences, is the mystery of innovation, its dynamical appearance and diffusion, frequency and coevolution, and how all of this links up with human cognition. Complex systems approaches offer promise to gain an integrated understanding of the many conflicting demands and forces which must be managed if our societies are to move toward sustainable development. In the context of globalization and the growing importance of long-distance interactions through a variety of networks, complex systems analysis (including direct observations and simulation experiments) can help us explore a variety of issues related to economic development, social cohesion, or the environment at different geographical scales.

Finally, the rapidly growing influence of information and communication technologies in our societies and the large number of decentralized networks relying on these new technologies require research and management ideas coming from complex systems science. In particular, the trend in information science is moving from processors to networks, and this leads to the emergence of so-called "ubiquitous intelligence," a phenomenon that will play an increasing role in determining how the networks of the future will be designed and managed.

# 1. Questions

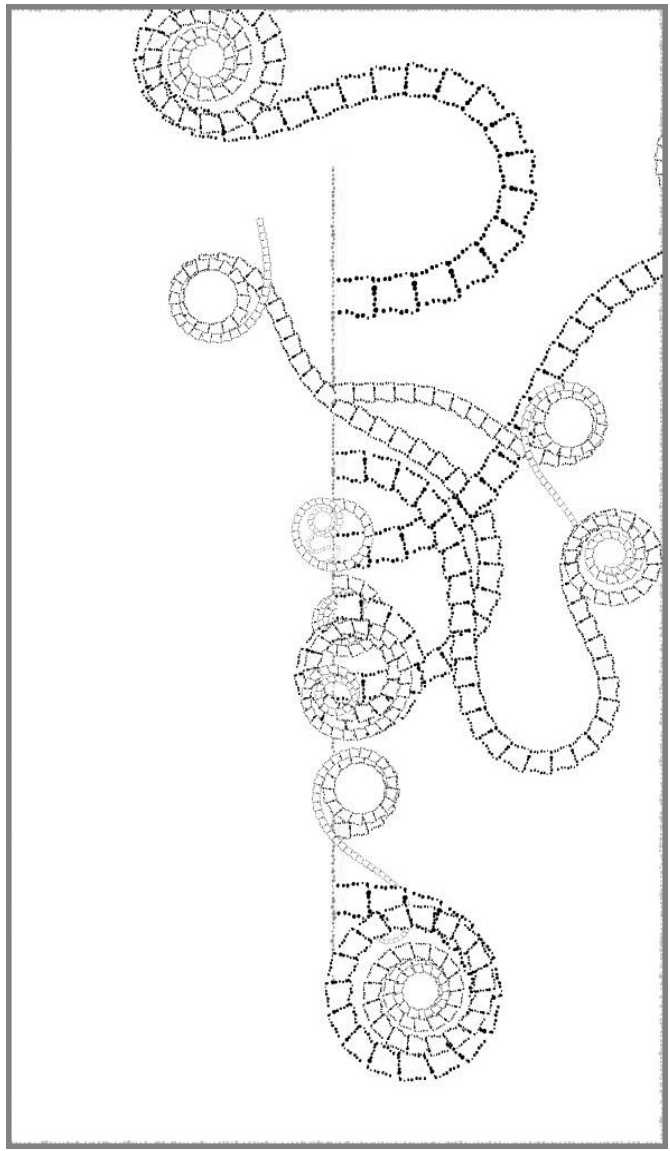

# 1.1. Formal epistemology, experimentation, machine learning

**Reporter:** Nicolas Brodu (INRIA – Rennes)

**Contributors:** Paul Bourgine (CREA, Ecole Polytechnique), Nicolas Brodu (INRIA – Rennes), Guillaume Deffuant (CEMAGREF), Zoi Kapoula (CNRS), Jean-Pierre Müller (CIRAD), Nadine Peyreiras (CNRS).

**Keywords:** Methodology, tools, computer, experimentation, modeling, validation, machine learning, epistemology, visualization, interaction, functional entity, formalization, phenomenological reconstruction.

## Introduction

The modern world, especially in medicine, in the social sphere and in the environment, increasingly depends on or confronts very large systems comprised of many interacting entities. The data collected on such systems, typically on enormous scales, poses formidable challenges for efforts to reconstruct their multiscale dynamics and their multiple downward and upward influences. The task requires the help of a formal epistemology and massive computation, and a generalization of the kind of "open science" originally inspired by the high-energy physics community.

The task of understanding a phenomenon amounts to finding a reasonably precise and concise approximation for its structure and behaviour, which can be grasped by the human mind. As it is, human intuition unaided cannot handle the intrinsic subtleties and non-intuitive properties of complex systems. Ideally, optimal formal techniques may provide us with candidate concepts and relations, which can then serve as a basis for the human experimental work. When the optimal forms found by the theory do not match the optimal concepts for the human brain, the reason for this discrepancy will itself be the subject of further investigation. Understanding complex systems thus requires defining and implementing a specific formal and applied epistemology. New methods and tools have to be developed to assist experimental design and interpretation for:
- Identifying relevant entities at a given time and space scale.
- Characterizing interactions between entities.
- Assessing and formalizing the system behaviour.

The strategy from experimental design to post hoc data analysis should reconcile the hypothesis- and data-driven approaches by:
- Defining protocols to produce data adequate for the reconstruction of multiscale dynamics.
- Bootstrapping through the simultaneous building of a theoretical framework for further prediction and experimental falsification.
- A functional approach at different levels, leading to the construction of adequate formalisms at these levels. There is no theoretical guarantee that one formal level could then be deducible in any way from another, but this does not matter:

Phenomenological reconstruction steps are preferable at each relevant level for the comprehension of the system.

The methodology begins with observation and data collection. However, there comes a point at which it is not relevant to go on collecting data without knowing whether they are really required for understanding the system behaviour. Phenomenological reconstruction leads to data parameterisation and obtained measurements should allow further detection and tracking of transient and recurrent patterns. These features themselves only make sense if they are integrated into a model aiming to validate hypotheses. We expect here to find a model consistent with the observations. The mere fact of building the model necessitates the formalization of hypotheses on the system behaviour and underlying processes. Part of the understanding comes from there, and more comes from the possibility of validating the model's predictions through experimentation. This last point is depicted on the right-hand side of the graph below.

*Formal & Applied Epistemology*

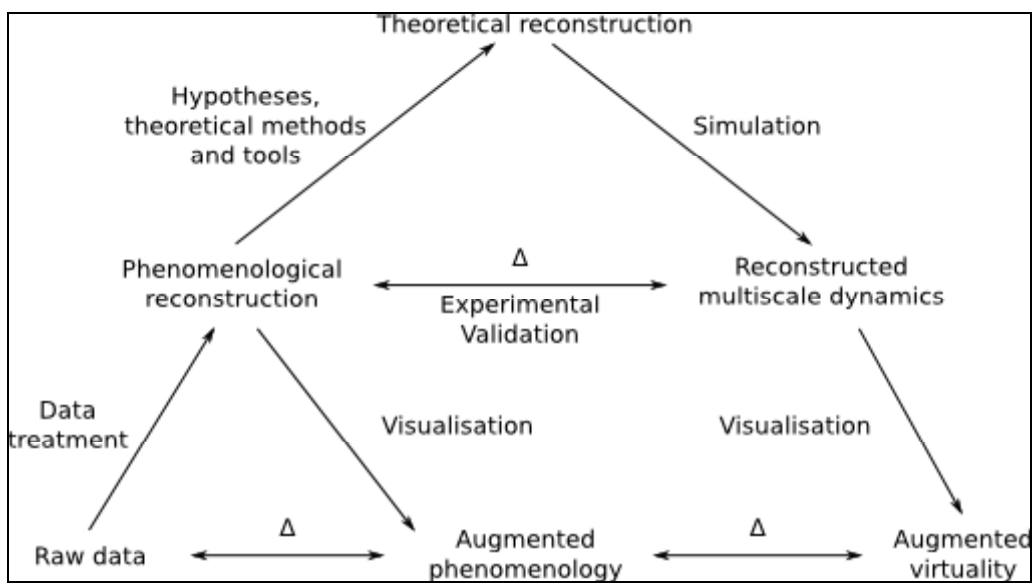

*The workflow of theoretical reconstruction*

The integration of computer science is an essential component of this epistemology. Computer science should provide:
- Exploratory tools for a data-based approach. Unsupervised machine learning can provide people with candidate patterns and relations that unaided human intuition would never detect. Active machine learning is concerned with determining which experiments are best suited to test a model, which is at the heart of the above epistemology.
- Tools for comparison between the model (hypothesis-driven) and the observations. Supervised learning corresponds to exploring the model parameter space for a good fit to the data. Auto-supervised learning is used when a temporal aspect allows the continuous correction of model predictions with the observed data corresponding to these predictions.

Computer science methods and tools are required in the following steps:
- Human-machine interactions: visualization and interaction with data, ontologies and simulations.

- Building ontologies of relevant functional entities at different levels.
- Constructing hypotheses, formalizing relations between entities, designing models.
- Validating the models.

We expect certain fundamental properties from computer science methods and tools:
- Generic tools should be as independent as possible from a logical (interpretation) framework. In particular, because of the varying cultural habits of different disciplines and the specificities of each system, it is preferable to propose a collection of independent and adaptable tools, rather than an integrated environment that would not cover all cases anyway.
- Independence should also apply for the software aspect (for the usage and the evolution and the adaptation of the tools to specific needs). This requires free/libre software as a necessary but not sufficient condition.
- Tools need to be useful for a specialist and also usable by non-specialist. This can be achieved, for example, by providing domain-specific features with added value for the specialist as extensions (modules, etc.) of the generic tools.
- Readiness for use: The preconditions for the application of the tool should be minimal, the tool should not require a large engineering effort before it can be used.

**Main Challenges**
1. Computer tools for exploration and formalization
2. Computer assisted human interactions

## 1.1.1. Computer tools for exploration and formalization

The computer must be identified as an exploration and formalization tool and integrated into an epistemology of complex systems.

Some research domains currently embrace this approach and their efforts need to be furthered. Computational mechanics and its causal state reconstruction is one candidate technique which could possibly automate phenomenological reconstruction, but there are challenges concerning its real applicability. For example, we face obstacles in finding a practical algorithm for the continuous case, or in building significant statistical distributions with a limited number of samples (relative to the search space). Statistical complexity can also be considered as a useful exploratory filter to identify the promising zones and interacting entities in the system. Another research domain that could be integrated into the epistemology is the quantification of the generalization capabilities of learning systems (e.g. Vapnik et al.). Automated selection of the most promising hypotheses and/or data instances is the topic of active learning. Its application is particularly straightforward for exploring the behaviour of dynamical computer models, but more challenging for multiscale complex systems. The problem may be, for instance, to determine response surfaces which lead to a major change of behaviour (the collapse of an ecosystem, for instance). When the system is of high dimension, the search space is huge and finding the most informative experiments becomes crucial. Some analysis techniques are inherently multiscale (e.g. fractal/multifractal formalisms) and would need to be integrated as well. Dynamical regimes are a essential part of complex systems, where sustained non-stationary and/or transient phenomena maintain the state out of static equilibrium. Some of the existing mathematical and algorithmic tools should be adapted to this dynamic context and new ones may have to be created specifically. Research is also needed on how to integrate these dynamical aspects directly into the experimental and formal aspects of the above epistemology.

## 1.1.2. Computer assisted human interactions

The computer has become a necessary component of scientific epistemology, as an extension to the human experimentalist. Three kinds of interactions involving humans and machines might be considered:

- **Machine to human:** The human sensory system (sight, hearing, etc.) is exceedingly powerful for some tasks, such as detecting patterns in an image, but quite poor for tasks like visualizing relations in large-dimensional spaces and graphs. Research is needed to explore how machines might provide people with an adequate representation of a complex system, in a form suitable for the human sensory system.
- **Human to machine:** The feedback and control that an unaided human can perform on a complex system is similarly limited. For example, when people act as the discriminating element for repeated decision-making (e.g. attributing/selecting fitness criteria of model parameters), the rate at which people can make decisions limits the execution speed of the algorithm. As a parallel to the visualization problem, human interaction capabilities on a large-dimensional simulation are relatively poor, especially with conventional devices such as a mouse and keyboard. Finding controls (software or hardware) adapted to the human morphology and limitations is another part of this human/complex system interaction challenge.
- **Human to human:** The computer should help human communication. For instance, knowledge from domain experts is often lost when non-specialist computer scientists formalize and create the experiments that experts need. Ideally, the computer should be a tool that enhances - not hampers - cross-disciplinary communication, and should be directly usable by the experts themselves for designing experiments and models and running simulations. But the use of the computer as a facilitator of human-to-human relations is not limited to interdisciplinary aspects. The computer should become an integral part of the collaborative process necessary to handle complex systems.

# 1.2. Stochastic and multiscale dynamics, instabilities and robustness

**Reporter:** Daniel Schertzer (Meteo France)

**Contributors:** Pierre Baudot (Inaf CNRS), Hughes Berry (INRIA), François Daviaud (CEA), Bérengère Dubrulle (CEA), Patrick Flandrin (CNRS ENS Lyon), Cedric Gaucherel (INRA), Michael Ghil (ENS Paris), Gabriel Lang (AGRPParis Tech), Eric Simonet (CNRS).

**Keywords:** Random dynamical systems, non stationarity, long range/ short range dependence, local/nonlocal interactions, discrete/continuous scaling, cascades, wavelet/multifractal analysis, multiscale modeling and aggregation/disaggregation, pattern recognition, graph dynamics, extremes distribution and large deviations.

## Introduction

Hierarchical structures extending over a wide range of space-time scales are ubiquitous in the geosciences, the environment, physics, biology and socio-economic networks. They are the fundamental structures building up our four-dimensional world's complexity. Scale invariance, or "scaling" for short, is a powerful mathematical tool for characterising these structures and inferring properties across scales, instead of dealing with scale-dependent properties. Whereas scaling in time or in space has been investigated in many domains, four-dimensional scaling analysis and modeling are still relatively little used and under-developed, even though it is indispensable for describing, estimating, understanding, simulating and predicting the underlying dynamics. Rather complementary to this approach, random dynamical system theory is also a powerful approach for grasping multiscale dynamics. This theory is likely to provide interesting generalizations of what we have learned from deterministic dynamical systems, particularly in the case of bifurcations. Other important domains of investigation are phase transitions, emerging patterns and behaviours which result when we move up in scale in the complex four-dimensional fields.

> **Main Challenges**
> 1. The cascade paradigm
> 2. Random dynamical systems and stochastic bifurcations
> 3. Phase transitions, emerging patterns and behaviour
> 4. Space-time scaling in physics and biology

## 1.2.1. The cascade paradigm

The idea of structures nested within larger structures, themselves nested within larger structures and so on over a given range of space-time scales has been in physics for some time, and could be traced back to Richardson's book (Weather Prediction by Numerical Processes, 1922) with his humoristic presentation of the paradigm of cascades. This paradigm became widely used well beyond its original framework of atmospheric turbulence, in such fields as ecology, financial physics or high-energy physics. In a generic manner, a cascade process can be understood as a space-time hierarchy of structures, where interactions with a mother structure are similar in a given manner to those with its daughter structures. This rather corresponds to a cornerstone of multiscale stochastic physics, as well as of complex systems: a system made of its own replicas at different scales.

Cascade models have gradually become well-defined, especially in a scaling framework, i.e. when daughter interactions are a rescaled version of mother ones. A series of exact or rigorous results have been obtained in this framework. This provides a powerful multifractal toolbox to understand, analyse and simulate extremely variable fields over a wide range of scales, instead of simply at a given scale. Multifractal refers to the fact that these fields can be understood as an embedded infinite hierarchy of fractals, e.g. those supporting field values exceeding a given threshold. These techniques have been applied in many disciplines with apparent success.

However, a number of questions about cascade processes remain open. They include: universality classes, generalized notions of scale, extreme values, predictability and more generally their connection with dynamical systems either deterministic-like (e.g. Navier-Stokes equations) or random (those discussed in the next section). It is certainly important to look closely for their connections with phase transitions, emerging patterns and behaviours that are discussed in the corresponding section. Particular emphasis should be placed on space-time analysis and/or simulations, as discussed in the last section on the general question of space-time scaling.

## 1.2.2. Random dynamical systems and stochastic bifurcations

Along with mathematicians' interest in the effects of noise on dynamical systems, physicists have also paid increasing attention to noise effects in the laboratory and in models. The influence of noise on long-term dynamics often has puzzling non-local effects, and no general theory exists at the present time. In this context, L. Arnold and his "Bremen group" have introduced a highly novel and promising approach. Starting in the late 1980s, this group developed new concepts and tools that deal with very general dynamical systems coupled with stochastic processes. The rapidly growing field of random dynamical systems (RDS) provides key geometrical concepts that are clearly appropriate and useful in the context of stochastic modeling.

This geometrically-oriented approach uses ergodic and measure theory in an ingenious manner. Instead of dealing with a phase space S, it extends this notion to a probability bundle, S x probability space, where each fiber represents a realization of the noise. This external noise is parametrized by time through the so-called measure-preserving driving system. This driving system simply "glues" the fibers together so that a genuine notion of flow (cocycle) can be defined. One of the difficulties, even in the case of (deterministic) nonautonomous forcing, is that it is no longer possible to define unambiguously a time-independent forward attractor. This difficulty is overcome using the notion of pullback attractors. Pullback attraction corresponds to the idea that measurements are performed at present time t in an experiment that was started at some time s<t in the remote past, and so we can look at the "attracting invariant state" at time t. These well-defined geometrical objects can be generalized with randomness added to a system and are then called random attractors. Such a random invariant object represents the frozen statistics at time t when "enough" of the previous history is taken into account, and it evolves with time. In particular, it encodes dynamical phenomena related to synchronization and intermittency of random trajectories.

This recent theory presents several great mathematical challenges, and a more complete theory of stochastic bifurcations and normal forms is still under development. As a matter of fact, one can define two different notions of bifurcation. Firstly, there is the notion of P-bifurcation (P for phenomenological) where, roughly speaking, it corresponds to topological changes in the probability density function (PDF). Secondly, there is the notion of D-bifurcation (D for dynamical) where one considers a bifurcation in the Lyapunov spectrum associated with an invariant Markov measure. In other words, we look at a bifurcation of an invariant measure in a very similar way as we look at the stability of a fixed point in a deterministic autonomous dynamical system. D-bifurcations are indeed used to define the concept of stochastic robustness through the notion of stochastic equivalence. The two types of bifurcation may sometimes, but not always be related, and the link between the two is unclear at the present time. The theory of stochastic normal form is also considerably enriched compared to the deterministic one but is still incomplete and more difficult to establish. Needless to say, bifurcation theory might be applied to partial differential equations (PDEs) but even proving the existence of a random attractor may appear very difficult.

### 1.2.3. Phase transitions, emerging patterns and behaviour

Phase transitions are usually associated with the emergence of patterns and non-trivial collective behaviour, for instance due to the divergence of a correlation length. Beyond the classical example of glassy systems, these features have been recently observed in shear flows, where the transition from laminar to turbulence occurs discontinuously through gradual increasing of the Reynolds Number. In such a case, the order parameter is the volume fraction occupied by the turbulence as it slowly organizes into a band pattern, with a wavelength that is large with respect to any characteristic size of the system.

A similar transition seems to occur in cortical dynamics, when experimenters increase the forcing of the sensory flow, using spectral or informational measures as an order parameter. When subjected to simple visual input, neuronal processing is almost linear and population activity exhibits localized blob patterns. When subjected to more informational and realistic stimuli, the neuronal processing appears to be highly nonlinear, integrating input over large spatial scales (center-surround interaction) and population patterns become more complex and spatially distributed.

The present challenge is to build a simple stochastic model which can account for the emerging structures generated by the dynamic and their dependence on the forcing. A more fundamental long-term aim is to catch both glassy and turbulent flow dynamics under such formalism.

A novel approach consists in considering a population of agents which have their own dynamics and characterizing their collective behaviour at different observation scales through gradual aggregation.

The simplest way to aggregate agents is to sum an increasing number of them. When they are identically distributed and independent random variables, the law of large numbers and the central limit theorem apply and the resulting collective evolution is analogous to the individual one. The result does not change when the dependence is short range; this would be the equivalent of the laminar phase. As the spatial dependence becomes long range, the nature of the collective behaviour changes (lower rate of convergence, different limit process). By playing with the interaction range, one is therefore able to induce a phase transition.

Another kind of transition is observable if one allows for non-linear effects in the aggregation process. In such a case, the resulting process may be short-range or long-range dependent, even if the dynamics of the individual are simple (autoregressive short-range dependence in space and time).

A first task is to develop such aggregation methods for simple individual models and to investigate the joint effect of dependence and aggregation process. Examples of applications include geophysical problems, hydrology and hydrography, integrative biology and cognition.

### 1.2.4. Space-time scaling in physics and biology

*1.2.4.1. Empirical background*

Systems displaying a hierarchy of structures on a wide range of time and space scales occur almost everywhere in physics and biology.

In the geosciences, 'Stommel diagrams' displaying life time vs. size of structures (usually in log-log plots) span several orders of magnitude, but a satisfactory explanation of this state of affairs is missing.

In biology, metagenomics have recently been developed to explore microbial biodiversity and evolution by mining urban waste to improve our knowledge of the "tree of life," but the time structure is far from being reliably estimated.

In the area of computer and social networks, the web is the best-known example, but scale-invariant and small-world networks are encountered everywhere; in this case researchers have begun to explore the temporal aspects of such networks, but the connection between time evolution and spatial structure requires further attention.

*1.2.4.2. State of the art*

a) Taylor's hypothesis of frozen turbulence (1935), also used in hydrology, is presumably the simplest transformation of time scaling into space scaling. This is obtained by supposing that the system is advected with a characteristic velocity.

b) In other cases, the connection between space and time scaling is less evident. As already pointed out, this is the case for computer networks: (space) network topology and (time) computer traffic have been separately studied up to now. Morphogenesis is a research domain that requires the development of space-time scaling analysis.

c) More recently, the comparison of scaling in time vs. scaling in space has been used to determine a scaling time-space anisotropy exponent, also often called a dynamical exponent.

*1.2.4.3. What is at stake*

*a) Why do we need to achieve space-time analysis/modeling?*
Basically there is no way to understand dynamics without space and time. For instance, whereas earlier studies of chromosomes were performed only along 1D DNA positions, 4D scaling analysis is required to understand the connection between the chromosome structure and the transcription process.

*b) Data analysis*

We need to further develop methodologies:
- to perform joint time-space multiscale analysis either for exploratory analysis or for parameter and uncertainty estimations,
- to extract information from heterogeneous and scarce data,
- to carry out 4-D data assimilation taking better account of the multiscale variability of the observed fields,
- for dynamical models in data mining.

*c) Modeling and simulations*

We also need to further develop methodologies:
- to select the appropriate representation space (e.g. wavelets),
- to define parsimonious and efficient generators,
- to implement stochastic subgrid-scale parametrizations.

# 1.3. Collective behavior in homogeneous and heterogeneous systems

**Reporter:** Francesco Ginelli (Institut des Systèmes Complexes de Paris Ile-de-France )

**Contributors:** Cyrille Bertelle (Le Havre), Guillaume Beslon (LIRIS, Lyon), David Chavalarias (CREA – ISC-PIF, Paris), Valérie Dagrain (France), François Daviaud (CEA, Saclay), Jean-Paul Delahaye (Lille), Silvia De Monte (CNRS, Paris), Cédric Gaucherel (INRA), Jean-Louis Giavitto (IBISC, Evry), Francesco Ginelli (ISC Paris and CEA/Saclay), Christophe Lavelle (Institut Curie, Paris), André Le Bivic (CNRS SDV, Marseille), Jean-Pierre Müller (CIRAD), Francesco d'Ovidio (ENS Paris), Nadine Peyrieras (CNRS, Gif s/ Yvette), Eric Sanchis (Université Paul Sabatier), Fabien Tarissan (Ecole Polytechnique).

**Keywords:** Collective dynamics, population diversity, agent-based models, environment heterogeneity, stochastic partial differential equations, reconstruction techniques, mesoscopic description, Lyapunov analysis, phase synchronization.

## Introduction

From genetic and social networks to the ecosphere, we face systems composed of many distinct units which display collective behaviour on space and time scales clearly separated from those of individual units. Among many others, we can mention cellular movements in tissue formation, flock dynamics, social and economic behaviour in human societies, or speciation in evolution.

The complexity of such phenomena manifests itself in the non-trivial properties of the collective dynamics - emerging at the global, population level - with respect to the microscopic level dynamics. Many answers and insights into such phenomena can and have been obtained by analysing them through the lens of non-linear dynamics and out-of-equilibrium statistical physics. In this framework, the microscopic level is often assumed to consist of identical units. Heterogeneity is, however, present to varying extents in both real and synthetic populations. Therefore, the existing descriptions also need to encompass variability in both the individual units and the surrounding environment, and to describe the structures that emerge at the population level. Similarly, while a homogeneous environment (medium) is a useful approximation for studying collective dynamics, hardly any real environment, either natural or artificial, is homogeneous, and heterogeneity deeply influencing the structures, dynamics and fates of a population. The variability of the environment applies both on spatial and temporal scales. Examples include filaments and vortices in fluid media, patches and corridors in landscapes, or fluctuating resources.

From a methodological point of view, successful modeling of such influences requires, at least: the quantification of environmental heterogeneities at different scales; the improvement of the formalization of heterogeneity; the identification of the heterogeneity features that are relevant at the population level and the study of population responses to changes in these heterogeneities.

Also of crucial importance for our understanding of biological processes is an understanding of what generates heterogeneity, and how it influences the further development of emergent patterns at many different scales. During the early stages of embryogenesis in metazoans, for example, cell diversity -- which is required for further functional differentiation -- is generated from the non-homogeneous distribution of sub-cellular components, cell division and cell environment interaction. Models need to link the collective behaviour of cell populations, which underlie pattern formation, to cell diversification and differentiation. Both theoretical and experimental aspects of these questions have been almost completely unexplored so far. Finding how molecular and cellular behaviours are coupled in these processes is a main challenge of developmental biology.

Close interaction between physicists and biologists skilled in non-linear analysis, social scientists and computer scientists has proved to be a key ingredient for advances in handling these subjects.

> **Main Challenges**
> 1. Collective dynamics of homogeneous and/or heterogeneous units
> 2. Collective dynamics in heterogeneous environments
> 3. Emergence of heterogeneity and differentiation processes, dynamical heterogeneity, information diffusion

## 1.3.1. Collective dynamics of homogeneous and/or heterogeneous units

In the past few years, researchers have devoted considerable effort to studying and characterising the emergence of collective phenomena through observation, experiment and theory. Research has explored a wide range of systems, from nano-structures and granular matter to neuronal dynamics and social organization in the animal kingdom (including human societies).

The intrinsic dynamical nature of these phenomena bears marked similarities with the physics of non-linear systems. Studies have indeed documented a number of dynamical paths to organised collective behaviour having strong resonance with systems in physics: phase synchronization in interacting oscillating systems, ordering phase transition in systems of self-propelled agents, and self-organization and pattern formation in spatially-extended systems (e.g. ecological systems).

However, we are far from fully understanding the relation between microscopic dynamics and macroscopic properties. For instance, the emergence at the global level of non-trivial coherent dynamics out of unlocked microscopic oscillators, characterized by time scales much shorter than the macroscopic one, still lacks a general theoretical framework. While some researchers speculate that transport coefficients may be extracted from the long wavelength components of microscopic linear analysis (Lyapunov analysis), no clear connection has been established so far. Systems of self-propelled units seem to display anomalously large fluctuations in number density – unknown in ordinary equilibrium matter, though observed experimentally in granular media – but current theoretical models only partially account for such phenomena.

New insights are expected from the intermediate-scale mesoscopic description which bridge the microscopic and macroscopic levels by coarse-graining relevant quantities over appropriate local scales of space and time. Due to the importance of fluctuations in out-of-equilibrium phenomena, the resulting partial differential equations (PDEs) are expected to yield stochastic terms, often multiplicative in the coarse-grained fields. The analysis of such stochastic PDEs is an open challenge for physicists and mathematicians alike, both from the numerical and the analytical point of view, but powerful new techniques, such as the non-perturbative renormalization group, promise to shed new light on this subject in the near future.

Although researchers have so far focussed on systems made up of largely identical units, numerous problems of interest involve systems, such as living organisms containing diverse cell types or ecosystems of many species composed of units of many different kinds. The task of fully understanding the emergence of collective phenomena in such systems requires taking into account the interactions among such elements, and poses numerous questions. To what extent can heterogeneous systems be reduced to homogeneous ones? In other words, is a wide degree of heterogeneity an irreducible feature of certain systems (for instance, complex ecological niches), and simply beyond any description in terms of simpler, decoupled few-species models? Are the emergent properties of homogeneous systems conserved in heterogeneous ones, and what are the specific features that arise at the collective level from microscopic heterogeneity? How do new emergent properties relate to previous results obtained in a more homogeneous context? Can the theoretical results for homogeneous systems be extended towards heterogeneous systems? Can we extend tools already developed for modeling collective dynamics to take heterogeneity into account (agent-based simulation, for example, can be very naturally extended) or do we have to develop specific new tools?

At the theoretical level, a number of avenues have shown promise. The study of simple systems composed of coupled oscillators with heterogeneous frequencies, for example, may open new insights into more practical systems, while the important role played by synaptic plasticity in neuronal dynamics has long been recognized.

Segregation between different species can be readily described using heterogeneous agent-based models, while cells can be seen as an inhomogeneous fluid. Thus, theoretical results about the behaviour of such systems (e.g. phase transitions, diffusion in crowed heterogeneous media, etc.) could shed new light on many open questions in molecular and cellular biology, such as the organization of the cell nucleus, diffusion in membranes, signal transduction, or the regulation of transcription.

Finally, it is worth recalling that the theoretical approach must be developed in parallel with experimental observations. Model studies need to provide results in a form which can be compared and validated with quantitative experiments. In particular, spatial reconstruction techniques – allowing to measure the three-dimensional position and trajectory of each unit inside a large group - are proving increasingly useful for extracting information at the level of microscopic dynamics.

### 1.3.2. Collective dynamics in heterogeneous environments

The complexity of collective dynamics reflects the properties of individual units, the interactions between them, but also influences coming from the surrounding environment. Assessing how environmental heterogeneity influences collective systems poses a central challenge in many fields, including biology, geosciences, computer and social sciences.

The complex systems approach should provide a unifying framework for investigating the effect of environmental heterogeneity on population dynamics. In particular, progress is needed in the following directions:

- Multiscale analyses: the observation and measurement of environmental heterogeneity requires new tools for its detection against noisy backgrounds and its analysis at multiple scales. In landscape ecology, for example, researchers need to capture the scaling sensitivity of mosaic heterogeneity, yet still lack technical tools for this purpose. A similar problem arises in plankton studies: turbulence gives spatial and temporal structure to populations on scales ranging from centimetres to the entire oceanic basin, and from minutes to years, but observations currently cover only portions of this range.
- Formalization: the capture of heterogeneity within models requires novel means for mathematical representation. Equations, algorithms and geometric representations must encompass environmental heterogeneity at different scales and describe and couple the environment with the underlying dynamics of individual system units. For example, long-range hydrodynamical interactions should be included in models describing the collective motion of bacteria swimming in viscous fluids. Evolution of the vegetal cover has been formalized using differential equations for continuous diffusion processes or percolation-based approaches. Yet a mathematical formalization of more discontinuous environments, either in terms of environment heterogeneity or of constituting units, remains to be achieved.
- Identification of key environmental features: models cannot include a description of all possible sources of heterogeneity, and it is therefore important to identify the aspects of heterogeneity that are most relevant for the chosen description of the system. Heterogeneity can be examined in terms of information, texture, correlation parameters, or coherent structures selected for the collective dynamics under study. For example, landscape structures may exhibit different heterogeneity, depending on the properties influencing the collective dynamics: contrast often highlights barrier effects, while connectivity highlights preferential pathways in the mosaic. In a fluid, transport barriers and mixing regions organize the spatial distribution of tracers; nonlinear methods make it possible to extract such structures from the velocity field and to shed light on the interaction between turbulence and biochemical tracers.
- Changing environments: heterogeneity is often not defined once and for all, but can change over the course of time. Such changes can occur on time scales faster than those of the collective dynamics, or manifest themselves as slow drifts in the environmental properties. Both kinds of change affect microbiological populations, for example, living in environments where food availability and temperature are subject to intense fluctuations. An adequate description of adaptation and evolution of collective behaviour demands that such fluctuations be taken into account. When the population itself induces environmental modifications, it is the feedback between collective behaviour and environmental heterogeneity which shapes the coupled population-environment dynamics, as in the case of the biota-earth interaction in the wake of climate change.

### 1.3.3. Emergence of heterogeneity and differentiation processes, dynamical heterogeneity, information diffusion

From genetic networks to social networks and ecospheres, many natural systems display endogenous heterogeneity: heterogeneity that emerges from the very functioning of the system. Mechanisms producing such heterogeneity include cell differentiation in ontogeny, social and economic differentiation in human societies and speciation in evolution. The origin and role of this heterogeneity in the viability and maintenance of these large systems is still largely unknown. Yet its importance is recognized in the emergence of topological macro-structures which often underlie global functions. Understanding the emergence of heterogeneity and its maintenance is thus a challenge for our efforts manage and, possibly, control complex systems.

From a simple homogeneous structure (multiple copies of the same object or uniform topological space) there are four main types of process through which heterogeneity emerges, processes which can be classified in terms of both their Kolmogorov complexity and their logical depth (a measure of "organizational complexity" introduced by Charles Bennett).

- Random emergence: noise upon a regular simple structure (random perturbation). One observes an increase in Kolmogorov complexity, but no increase in organizational complexity.
- Coordinated or strongly constrained evolution. Example: the duplication of a gene gives two genes, allowing the divergence of their function, or the differentiation of individuals in a social structure (specialization, new functions, etc.). It is not necessarily associated with a significant increase in Kolmogorov complexity, but with an increase in organizational complexity ("crystallization of a computation").
- Mixed emergence: randomness and constraints play a role in the dynamical process of emergence. Examples: whole molecular and genetic modules are re-used and evolve, leading to morphogenetic and functional diversity; speciation by isolation and adaptation to various geographical constraints; several copies of an entity subjected to various conditions diverge by learning or mutual adjustment. In this case, there is an increase in both Kolmogorov complexity and organizational complexity.
- Emergence by "computation/expression of a pre-existing program". If the "computation" is fast and non-random, there is no increase in Kolmogorov complexity, nor in organizational complexity.

# 1.4. From optimal control to multiscale governance

**Reporter:** Jean-Pierre Müller (CIRAD)

**Contributors:** Frédéric Amblard (Université de Toulouse), Jean-Christophe Aude (CEA), David Chavalarias (Institut des Systèmes Complexes de Paris Ile-de-France / CNRS), Jean-Philippe Cointet (Institut des Systèmes Complexes de Paris Ile-de-France / INRA ), Valérie Dagrain (Consultante technologie information), Guillaume Deffuant (Meteo-France & U. Paris-Est Cemagref), Sarah Franceschelli (Paris Jussieu), Pablo Jensen (ENS Lyon), Maud Loireau (IRD), Denise Pumain (Université Paris 1), Ovidiu Radulescu (Université de Rennes 1), Eric Sanchis (Université Paul Sabatier).

**Keywords:** Governance, control, multi-criteria, optimal control, viability, negotiation, multi-level, exploration/exploitation compromise, uncertainty, social acceptability, participation.

## Introduction

In acting on a complex system, institutions in charge of its governance first face the problem of defining desired objectives. Often, these objectives must integrate the conflicting interests and points of view of diverse stakeholders at multiple scales. Then, in order to compromise and to decide on policy actions to match the objectives, it is necessary to build an appropriate understanding of the phenomena, often through modeling which includes the effect of the potential actions. (Here, we touch again on the general problem of modeling and reconstructing dynamics from data, addressed in another part of the roadmap). Unfortunately, current methods (reinforcement learning, viability, etc.) for assessing policy actions only work practically for models in state spaces of low dimensionality. Progress can be sought in two directions: either by extending these methods to multiscale and higher dimensionality dynamics and multi-level actions (e.g. central and decentralized), or by projecting multiscale dynamics into smaller spaces. The use of stylized dynamics, when possible, is another research direction that could open new possibilities for managing good policy actions on complex dynamics. Finally, dynamics are often uncertain and partially unknown, which implies a difficult compromise between exploitation of the better known parts of the dynamics and exploration of worse known parts. This problem can be extended to the reformulation of the problem (including the objectives). This framework similarly addresses problems of control and design.

> **Main Challenges**
> 1. Extending the scope of optimal control
> 2. Projecting complex dynamics into spaces of smaller dimension
> 3. Projecting optimal control into high and multiscale dimension space
> 4. Extending exploration / exploitation compromise to problem reformulation
> 5. Co-adaptation of governance and stakeholders' objectives

## 1.4.1. Extending the scope of optimal control

Current methods of optimal control can deal with uncertain non-linear dynamics, and with flexible definitions of the objectives (in viability theory, for instance), but they are limited by the curse of dimensionality: these methods must sample the state space with a given precision, and this requires computational power which increases exponentially with the dimensionality of the state space. Extending these methods to spaces of larger dimensions is therefore crucial to enable their use in the context of complex systems.

One potential approach for addressing these questions is to develop weaker frameworks than optimal control. For instance, control may seek resilience and viability, or the maintenance of some important functional properties, without demanding the traditional objective of optimal control, which is to maximise a function.

Finally, in some cases, mixing mathematical optimisation of action policies and participatory approaches within an iterative dialogue could provide a good compromise between flexibility, social acceptability and rationality. Such approaches would require a specific methodological focus on how to define parts of the problem which can be treated automatically, and how to integrate the results of these optimising algorithms efficiently with other aspects of a group decision process.

## 1.4.2. Projecting complex dynamics into spaces of smaller dimension

Another possibility to tackle the limits of current methods for control is to reduce the dimensionality of complex dynamics (for instance, through the identification of slower parts of the dynamics, the aggregation of the state space, the definition of stylized dynamics and so on). This type of work is also very important in negotiation and formulation processes, in order to give stakeholders intelligible materials from which they can easily express their views. We do not know of reduction approaches directed at the local views of the different stakeholders: such approaches would be very interesting.

Dimensionality reduction applies to both data (information) and models. Statistical techniques based on Principal Component Analysis determine a linear space containing the essential information. They do not apply to non-linear correlation, when projection should lead to curved manifolds. New methods are needed to cover this case as well. Non-linear Independent Components represent one possible direction of research. Classical model reduction techniques, such as averaging, singular perturbations, or calculation of invariant manifolds, are based on separation of time and space scales. These methods are currently used for applications in physics and chemistry and they should be adapted to take into account the specificities of other domains. Furthermore, complex models are only partially specified. For instance, models in biology are qualitative and knowledge of parameters is only partial. Classical model reduction methods start from models that are completely specified (all parameters are known). There is a need for model reduction techniques which can replace numerical information by ordinal information (one parameter is much smaller than others) or other types of qualitative information.

## 1.4.3. Projecting optimal control into high and multiscale dimensional space

Another possibility is to extend optimal control (and any developments beyond optimal control) to high dimensional, multiscale systems. Such an extension should consider the possibility of using distributed actions at different levels, particularly in a decentralized way. This is a challenging aim, even if the effect of the controls is perfectly known, because not only the system but the control as well is multi-dimensional, with potentially non-linear effects of control coordination. Research requires new approaches to advance in this direction.

The scope of this approach might also be extended to cases with multiple objectives at different scales. Such a proposition involves introducing the concept of a "complex objective," and would probably require new formalisms to describe the architecture and links between these multiscale objectives. Since they are described at different levels, current control methods are not suitable for tackling this concept. New research should therefore be undertaken in this field using either centralized or distributed control. The latter method is appealing since it allows different semantics of control and actions at different scales. This concept raises several questions, including: how to couple and synchronise controllers; how to deal with simultaneous and opposite actions on the system; how to handle the different hierarchical levels; how to mix participation/decision making/optimisation; how to implement distributed control with a single global objective or multiple local objectives or both.

## 1.4.4. Extending the exploration/exploitation trade-off to governance analysis

Decision-makers often have multiple possibilities for action, and from these have to choose how to allocate their resources. The eventual outcome of policy actions, relative to the objectives, often remain imperfectly known, making policy evaluation very difficult. As a consequence, decision makers regularly face a trade-off between further exploration of the different available opportunities for action, and exploitation of certain selected opportunities. Exploring opportunities requires experiments at appropriate scales in time and space, and therefore the expenditure of resources. These expenses must be compared with the potential benefit of such exploration, compared with the mere exploitation of known routines.

In the framework of governance, exploration is necessarily made at a given scale of time and space, whereas governance initiatives are performed within open systems and therefore at several scales of space and time. The challenge is thus to propose methods and tools which can go beyond constraints of exploration and bridge the gap between the results of exploratory experiments and full-scale *in vivo* implementation of governance actions. These methods would have to take into account the reactive and adaptable nature of the targeted systems, as specified in challenge 1.

## 1.4.5. Co-adaptation of governance and stakeholders' objectives

In a multi-level context, identifying the stakeholders and territory concerned is a problem in itself.

The co-existence of different objectives, which may be in conflict, raises problems for the management or regulation of the system. Moreover, in some circumstances, the fact that these objectives may evolve with the environment (social context) or may adapt to a dynamical context (Ambient Intelligence) makes the system even more complex to manage or design.
We can focus on two challenges**:**

*1.4.5.1. The static dimension: governance in the context of heterogeneity of stakeholders, their points of view and interests*

The challenge is to develop models and methodologies to take into account the large heterogeneity of stakeholders' viewpoints and interests, which is reinforced by the entanglement of a large range of space and time scales. Multi-criteria analysis is a starting point for solving such problems, but it must be extended to incorporate several parallel objectives, and to include the reformulation process. Moreover, the choice of indicators linked to given objectives or their achievement must include stakeholder participation and be easy to use. On the other hand, the theoretical consequences of the choice of indicators, and particularly the potential biases they may introduce, must be carefully investigated. These tools and methods should also help determine criteria for analysing the adequacy of the objectives (any-time evaluation) and progress towards achieving them**.**

*1.4.5.2. The dynamical dimension: evolution of stakeholders' objectives and viewpoints in the governance process*

The challenge is to develop models and methodologies to take into account the feedback loops associated with self-regulation mechanisms, as well as the interdependence of particular interests during the governance process. For example, changes in the process by which decision-makers and stakeholders interact may alter their conceptions of the objectives, and the problem itself, and this alteration may then come back to affect the interaction process. This process becomes even more complex in social settings, with efforts to coordinate multiple objectives at the collective level. The time scales of model formation, decision-making and the interaction process itself have to be taken into account.

These aspects of the problem deal with the question of governance, and focus on the participative context where co-learning becomes as important as collective negotiation and decision-making. Moreover, the results of the interaction during the governance process can lead to new views of the problem, and possibly new governance objectives (taking into account, for instance, social acceptability) or new structures in the multiscale architecture of the governance organizations.

# 1.5. Reconstruction of multiscale dynamics, emergence and immergence processes

**Contributors:** Paul Bourgine (Ecole Polytechnique), David Chavalarias (Institut des Systèmes Complexes de Paris Ile-de-France / CNRS), Jean-Philippe Cointet (Institut des Systèmes Complexes de Paris Ile-de-France / INRA ), Michel Morvan (ENS Lyon), Nadine Peyrieras (Institut de Neurobiologie CNRS).

**Keywords:** Micro-macro reconstruction, multi-level experimental protocol, emergence, immergence, dynamical systems, multiscale systems.

**Introduction**

The data collected from complex systems are often incomplete and difficult to exploit because they are limited to a single level, i.e. refer to observations made on particular scales of space and time. Gathering data effectively first requires the definition of common concepts and pertinent variables for models at each level. Another important problem is obtaining unified and coherent representations useful for integrating different levels of organization and for predicting the dynamics of the complete system. This goal can be achieved by defining pertinent variables at each level of organization, i.e. at different time (slow/fast) and spatial (macro/micro) scales, their relationships, and how they are coupled together in models that describe the dynamics at each level. The challenge is to make explicit and meaningful connections from micro to macro levels (emergence functions) and from micro to macro levels (immergence functions).

> **Main Challenges**
> 1. Building common and pertinent conceptual frameworks in the life sciences.
> 2. Achieving coherence in the modeling of complex systems.
> 3. Development of mathematical and computer formalisms for modeling multi-level and multiscale systems. Computer tools for exploration and formalization
> 4. Computer assisted human interactions

## 1.5.1. Building common and pertinent conceptual frameworks in the life sciences

The data collected from complex systems are often incomplete and therefore difficult to exploit. A main challenge is to find common methods to collect data at different levels of observation, which are coherent and compatible in the sense that they can be used in order to integrate behaviour a different levels of a multi-level (multiscale) system. Thus, it is necessary to find multiscale models which allow researchers to define pertinent experimental variables at each level and to achieve a common reference frame with data reproducibility in the different levels of organization of the complete system.

## 1.5.2. Achieving coherence in the modeling of complex systems

The goal is to find coherence in the definition of variables and models used at each level of the hierarchical system and to make compatible the models that are used to describe the dynamics at each hierarchical level of organization at given time and space scales.

As a first step, one must take natural constraints into account and verify fundamental laws at each level of description (definition of pertinent species, symmetry laws, physical laws, conservation laws and so on). The next step is to connect the description and models used at each level to those at other levels:
(i) Modeling the dynamics at microscopic levels can be useful for defining boundaries for global variables and even to obtain correct interpretations for global variables.
(ii) Modeling the dynamics at macroscopic levels can be helpful for defining local functions and variables governing microscopic dynamics.

## 1.5.3. Development of mathematical and computer formalisms for modeling multi-level and multiscale systems.

The complexity of natural and social systems stems from the existence of many levels of organization corresponding to different time and space scales. A major challenge of complex systems science is to develop mathematical formalisms and modeling methods able to capture complete system dynamics by integration of activity at many, often hierarchically organised, levels. This goal can be achieved by defining emergence and immergence functions and integrating intra-level (horizontal) and inter-level (vertical) couplings.

Mathematical models used to describe the dynamics of natural and social systems involve a large number of coupled variables for quantities at different scales of space and time. These models are in general nonlinear and difficult to handle analytically. Therefore, it is crucial to develop mathematical methods which allow one to build a reduced system governing a few global variables at a macroscopic level, i.e. at a slow time scales and long spatial scales. Among open questions, we mention the definition of pertinent variables at each level of organization. It is also necessary to obtain emergence (resp. immergence) functions that allow analysis to jump from a microscopic (resp. macroscopic) level to a macroscopic (resp. microscopic) level, to study the coupling between the different levels and therefore the effects of a change at one level of a hierarchy on the dynamics at others.

Methods based on the separation of time scales already allow the aggregation of variables and are used in mathematical modeling for integrating different hierarchical levels. However, such multi-level modeling methods need to be extended to computer modeling and particularly to IBM (Individual Based Models), and this constitutes a very promising research theme. Also, the comparison of multi-level models to experimental data obtained at different levels remains a major challenge to be investigated in parallel with the development of mathematical and computer modeling methodologies for multi-level systems.

# 1.6. Designing artificial complex systems

**Reporter**: René Doursat (Institut des Systèmes Complexes / CREA, Ecole Polytechnique)

**Contributors**: Jean-Christophe Aude (CEA), Sofiane Ben Amor (Eurocontrol), Marc Bui (EPHE), René Doursat (Institut des Systèmes Complexes / CREA, Ecole Polytechnique), Jean-François Mangin (CEA), Jean Sallantin (CNRS LIRMM).

**Keywords:** artificial assistants, virtual simulations, functional modeling and regulation, bio-inspiration, autonomous and evolutionary systems.

## Introduction

Modeling and simulation are crucial complementary tools in the exploration of complex systems. Striking advances in computer networks and high-performance calculation have stimulated the rapid development of complex systems research in many scientific fields, and strong interactions between disciplines. Information and communication technologies represent today a major tool of investigation in complex systems science, often replacing analytic and phenomenological approaches in the study of emergent behaviour. In return, information technologies also benefit from complex system research. Artificial complex systems can be created to analyse, model and regulate natural complex systems. Conversely, new and emergent technologies can find inspiration from natural complex systems, whether physical, biological or social.

> **Main Challenges**
> 1. Using artificial complex systems for the understanding and regulation of natural complex systems
> 2. Finding inspiration in natural complex systems for the design of artificial complex systems
> 3. Building hybrid complex systems.

## 1.6.1. Using artificial complex systems for the understanding and regulation of natural complex systems

Natural complex systems (NCS) include systems found in nature (natural patterns, biological organisms, the ecosphere, etc) but also systems spontaneously originating from human activity (cities, economies, transportation systems, etc.) A key application of artificial complex systems (ACS) is to assist the description, generation and support of these NCS. One major challenge is to design and develop systems capable of exploring NCS in a systematic way, or regulating such systems. In particular, ACS design can complement human collective intelligence by integrating different levels of expertise and harmonising or managing contradictions in collaborative works. Such artificial systems can be based on structures and principles of function quite different from the natural systems they observe. An ACS could serve to regulate, schedule, repair or modify the NCS. The execution of ACS can be asynchronous and separate from the NCS, or it can be integrated with it.

**Examples:**

- Reconstructing the topology of neural connections in the brain by means of neuro-imagery and artificial vision based on a distributed architecture
- Observation of interest groups and interaction networks on the Internet (forums, blogs, instant messaging) through software agents
- Airflight dynamics and network

## 1.6.2. Finding inspiration in natural complex systems for the design of artificial complex systems

In order to create technological systems that are autonomous, robust and adaptive, new engineering approaches must draw inspiration from NCS. For example, in computer security, systems able to mimic the biological immune system can provide useful solutions against continuously evolving attacks on computer networks. These ACS are built upon intrinsically distributed, self-organizing and evolutionary entities. They reproduce the original behaviour and organizational principles that are found in NCS but have no equivalent in traditional technical design. In some domains, biology could even replace physics at the foundation of new engineering principles.

NCS provide rich sources of ideas in the development of decentralized systems which can display robustness, modularity, and autonomy in dynamically changing environments (i.e., "ubiquitous computing", "ambient intelligence"). ACS should be able to reproduce the dual principles of cooperation and competition that are observed in NCS.

On the other hand, bio-inspired artificial design is not constrained by any fidelity to the original NCS. Computer and technological innovation can free designers from experimental data or real examples of functioning mechanisms. Examples include neural networks inspired by neuroscience and genetic algorithms by Darwinian evolution. ACS created this way can also play a heuristic exploratory role for NCS. Engineering inventions allow us to better understand, even predict the natural phenomena that inspired them.

**Examples:**
- Neuro-inspired artificial intelligence and robotics
- Collective optimization and swarm intelligence inspired from social animal behaviour
- Evolutionary robotics
- Intelligent materials, auto-assembling materials, and morphogenetic engineering (nanotechnologies)
- Ambient intelligence
- Computer security inspired by immune systems or social interactions

## 1.6.3. Design of Hybrid Complex Systems

The rapid dissemination of computing devices and systems in our society (cellphones, PDAs, etc.) and the intricacy and profusion of their interconnections constitute a major case of hybrid or "techno-social" complex systems. Such systems can be studied as complex communities combining natural and artificial agents. Users can instruct machines, themselves capable of autonomous learning and adaptation to their environment.

# 2. Topics

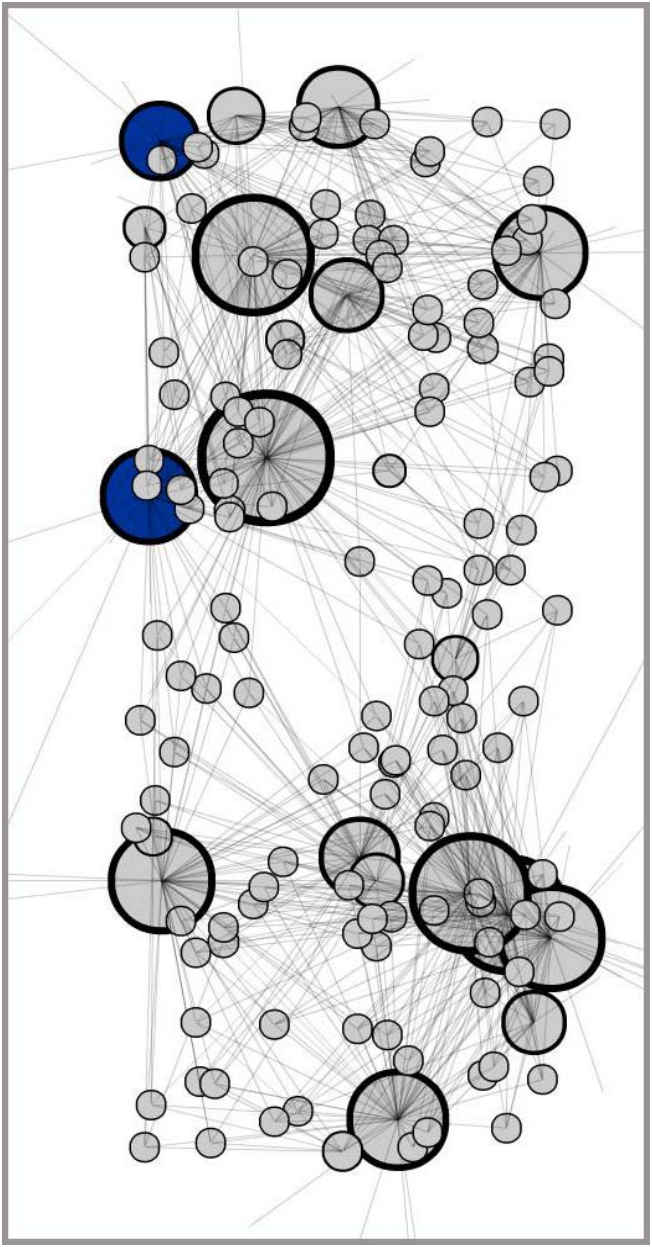

# 2.1. Complex matter

**Reporter:** François Daviaud (CEA)

**Contributors:** Giulio Biroli (CEA), Daniel Bonamy (CEA), Elisabeth Bouchaud (CEA), Olivier Dauchot (CEA, Commissariat à l'energie atomique), François Daviaud (CEA), Marc Dubois (CEA SPEC), Berengère Dubrulle (CEA), François Ladieu (CEA), Denis L'Hôte (CEA).

**Keywords:** Glassy dynamics, slow relaxations, frustration and disorder, collective behaviors, out-of-equilibrium and nonlinear systems, self-organization, turbulence, dynamo action, fracture.

**Introduction**

The field of complex and non-equilibrium systems is currently driven by a large body of new experiments and theoretical ideas in various branches of physics, from condensed matter physics up to ultra-cold atomic physics and biology. Beyond their apparent diversity, these systems share a common characteristic: the emergence of complex collective behaviours from the interaction of elementary components. Glassy dynamics, out-of-equilibrium systems, the emergence of self-organized or self-assembled structures, criticality, percolating systems, domain wall propagation and pinning of elastic walls, nonlinear systems, turbulence and fracture propagation are some subjects of complex matter that can be addressed only with the tools developed for the study of systems of interacting entities. Understanding these phenomena also requires the development of new theoretical methods in statistical physics and the design of new types of experiments.

> **Main Challenges**
> 1. Non-equilibrium statistical physics
> 2. Damage and fracture of heterogeneous materials
> 3. Glassy dynamics: glasses, spin glasses and granular media
> 4. Bifurcations in turbulence: from dynamo action to slow dynamics

### 2.1.1. Non-equilibrium statistical physics

The long lasting interest for non-equilibrium phenomena has recently experienced a noticeable revival, through the development of novel theoretical ideas (especially on the symmetries of non-equilibrium fluctuations) and new areas of applications, ranging from many examples in condensed matter physics to other branches of physics (heavy ion collisions, the early universe) and also to other sciences, including biology (manipulations of single molecules). Non-equilibrium phenomena also play an important part in many of the interdisciplinary applications of statistical physics (modeling the collective behavior of animals, or social and economic agents).

A physical system may be out of equilibrium for either of the following two reasons:

- **Slow dynamics**. The microscopic dynamics of the system are reversible, so that the system possesses an authentic equilibrium state. The dynamics of some of the degrees of freedom are however too slow for these variables to equilibrate within the duration of the experiment. The system is therefore in a slowly evolving non-equilibrium state for a very long time (forever in some model systems). The characteristic features of this regime of non-equilibrium relaxation, including the violation of the fluctuation dissipation theorem, have been the subject of intense activity over the last decade. These phenomena are commonly referred to as "aging" phenomena (see the part on glassy dynamics).
- **Driven dynamics**. The dynamics of the system are not reversible, usually because of some macroscopic driving caused by external forces. For instance, an electric field induces a non-zero current across the system, thereby destroying the reversibility of the underlying stochastic dynamics. The system reaches a non-equilibrium stationary state, where it stays forever. There are also systems (at least model systems) where the lack of reversibility lies entirely at the microscopic level, and does not rely on any macroscopic external driving. The paradigm of such a situation is the celebrated voter model.

One of the most salient advances of the last decade has been the discovery of a whole series of general results concerning the symmetries of spontaneous fluctuations in non-equilibrium states. These theorems, associated with names such as Gallavotti, Cohen, Evans and Jarzynski, have been applied and/or tested in many circumstances, both by theory and experiment.

Most recent efforts in this area have been devoted to interacting particle systems. This broad class of stochastic systems is commonly used to model a wide range of non-equilibrium phenomena (chemical reactions, ionic conduction, transport in biological systems, traffic and granular flows). Many interacting particle systems can be investigated by analytical methods, whereas some of them have even been solved exactly.

Although the usual formalism of equilibrium statistical physics does not apply to out-of-equilibrium systems, it is now well-known that many of the tools developed in equilibrium settings can also be used out-of-equilibrium. This is in particular the case for the framework of critical behaviour, where concepts such as scale invariance and finite-size scaling have provided (largely numerical) evidence for universality in non-equilibrium systems. It is possible to investigate systems in which the non-equilibrium character stems not from the presence of gradients imposed, for instance, by boundary reservoirs, but because of the breaking of micro-reversibility - that is to say, time-reversal invariance - at the level of the microscopic dynamics in the bulk.

A large part of the research activity on non-equilibrium statistical physics is also centred on the various phase transitions observed in many contexts. Indeed, many non-equilibrium situations can be mapped onto each other, revealing a degree of universality going well beyond the boundaries of any particular field: for example, self-organized criticality in stochastic (toy) sand piles has been shown to be equivalent to linear interface depinning on random media, as well as to a particular class of absorbing phase transitions in reaction-diffusion models. Another prominent example is the jamming transition which bridges the fields of granular media and glassy materials. It has been studied experimentally thanks to a model experiment consisting in a sheared layer of metallic disks. Synchronization and dynamical scaling are, likewise, very general phenomena which can be related to each other and to the general problem of understanding universality out of equilibrium.

## 2.1.2. Damage and fracture of heterogeneous materials

Understanding the interrelation between microstructure and mechanical properties has been one of the major goals of materials science over the past few decades. Quantitative predictive models are even more necessary when considering extreme conditions – in terms of temperature, environment or irradiation, for example – or long-time behaviour. While some properties, such as elastic moduli, are well approximated by the average of the properties of the various microstructural components, none of the properties related to fracture – elongation, stress to failure, fracture toughness – follow such an easy rule, mostly: (i) because of the high stress gradient in the vicinity of a crack tip, and (ii) because, as the more brittle elements of microstructure break first, one is dealing with extreme statistics. As a result, there is no way that a material can be replaced by an "effective equivalent" medium in the vicinity of a crack tip. This has several major consequences

*2.1.2.1. Size effects in material failure*

In brittle materials, for example, cracks initiate on the weakest elements of the microstructures. As a result, toughness and life-time display extreme statistics (Weibull law, Gumbel law), the understanding of which requires approaches based on nonlinear and statistical physics (percolation theory, random fuse models, etc.).

*2.1.2.2 Crack growth in heterogeneous materials*

Crack propagation is the fundamental mechanism leading to material failure. While continuum elastic theory allows the precise description of crack propagation in homogeneous brittle materials, we are still far from understanding the case of heterogeneous media. In such materials, crack growth often displays a jerky dynamics, with sudden jumps spanning over a broad range of length-scales. This is also suggested from the acoustic emission accompanying the failure of various materials and - at much larger scale - the seismic activity associated with earthquakes. This intermittent "crackling" dynamics cannot be captured by standard continuum theory. Furthermore, growing cracks create a structure of their own. Such roughness generation has been shown to exhibit universal morphological features, independent of both the material and the loading conditions, reminiscent of interface growth problems. This suggests that some approaches issued from statistical physics may succeed in describing the failure of heterogeneous materials. Let us finally add that the mechanisms become significantly more complex when the crack growth velocity increases and becomes comparable to the sound velocity, as in impact or fragmentation problems, for instance.

*2.1.2.3. Plastic deformation in glassy materials*

Because of high stress enhancement at crack tips, fracture is generally accompanied by irreversible deformations, even in the most brittle amorphous materials. While the physical origin of these irreversible deformations is now well understood in metallic materials, it remains mysterious in amorphous materials like oxide glasses, ceramics or polymers, where dislocations cannot be defined.

## 2.1.3. Glassy dynamics

*2.1.3.1. Glasses*

The physics of glasses concerns not only the glasses used in everyday life (silicates), but a whole set of physical systems such as molecular glasses, polymers, colloids, emulsions, foams, Coulomb glasses, dense assemblies of grains, etc. Understanding the formation of these amorphous systems, the so-called glass transition, and their out-of-equilibrium behaviour is a challenge which has resisted a substantial research effort in condensed matter physics over the last decades. This problem is of interest to several fields from statistical mechanics and soft matter to material sciences and biophysics. Several fundamental open questions emerge: is the freezing due to a true underlying phase transition, or is it a mere crossover with little universality in the driving mechanism? What is the physical mechanism responsible for the slowing down of the dynamics and glassiness? What is the origin of the aging, rejuvenation and memory effects? What are the common concepts that emerge to describe the various systems evoked above, and what remains specific to each of them?

Interestingly, however, evidence has mounted recently that the viscous slowing down of super-cooled liquids and other amorphous systems might be related to the existence of genuine phase transitions of a very singular nature. Contrary to usual phase transitions, the dynamics of glass-formers dramatically slows down with nearly no changes in structural properties. We are only just beginning to understand the nature of the amorphous long-range order that sets in at the glass transition, the analogies with spin-glasses and their physically observable consequences. One of the most interesting consequences of these ideas is the existence of dynamical heterogeneities (DH), which have been discovered to be (in the space-time domain) the counterpart of critical fluctuations in standard phase transitions. Intuitively, as the glass transition is approached, increasingly larger regions of the material have to move simultaneously to allow flow, leading to intermittent dynamics, both in space and in time. The existence of an underlying phase transition and of dynamical heterogeneities should significantly influence the rheological and aging behaviours of these materials, which are indeed quite different from those of simple liquids and solids. As a consequence, progress in the understanding of glassy dynamics should trigger several technological advances. An important example where the peculiar properties of glasses are used in technology is the stocking of nuclear waste.

From an experimental point of view, the major challenges for the future have been transformed not only because progress in the domain has led to radically new questions, but also because new experimental techniques now allow researchers to investigate physical systems at a microscopic scale. New challenges for the years to come are: i) To study the local dynamical properties in order to unveil which changes in the way molecules evolve and interact makes the dynamics glassy, in particular why the relaxation time of supercooled liquids increases by more than 14 orders of magnitude in a small temperature window; ii) To provide direct and quantitative evidence that glassy dynamics is (or is not) related to an underlying phase transition; iii) To study the nature of the dynamical heterogeneities (correlation between their size and their time evolution, fractal dimensions, etc.); iv) To investigate the nature of the out-of-equilibrium properties of glasses, such as violation of the fluctuation-dissipation theorem, intermittence, etc.

From a theoretical point of view, the major challenge is to construct and develop the correct microscopic theory of glassy dynamics. This will consist both in unveiling the underlying physical mechanisms that give rise to slow and glassy dynamics and in obtaining a quantitative theory that can be compared to experiments. The main focus will again be on local dynamic properties, their associated length-scale and their relation to the growing time scales and the global properties of glassy dynamics.

*2.1.3.2. Spin glasses*

The expression "spin glasses" was invented to describe certain metallic alloys of a non-magnetic metal with few, randomly substituted, magnetic impurities, where experimental evidence for a low temperature phase showed a non-periodic freezing of the magnetic moments with a very slow and strongly history-dependent response to external perturbations. Basic fundamental ingredients of spin glasses are disorder and frustration. The frustration consists in the fact that the energy of all the pairs of spins cannot be minimized simultaneously. The theoretical analysis of spin glasses lead to the celebrated Edwards-Anderson model: classical spins on the sites of a regular lattice with random interactions between nearest-neighbour spins. This has led to many developments over the years, and the concepts developed for this problem have found applications in many other fields, from structural glasses and granular media to problems in computer science (error correction codes, stochastic optimization, neural networks, etc.).

The program of developing a field theory of spin glasses is extremely hard, with steady, slow progress. The theory is not yet able to make precise predictions in three dimensions. Numerical simulations face several difficulties: we cannot equilibrate samples of more than a few thousand spins, the simulation must be repeated for a large number of disorder samples (due to non-self-averaging), and the finite size corrections decay very slowly.

Spin glasses also constitute an exceptionally convenient laboratory frame for experimental investigations of glassy dynamics. The dependence of their dynamical response on the waiting time (aging effect) is a widespread phenomenon observed in very different physical systems such as polymers and structural glasses, disordered dielectrics, colloids and gels, foams, friction contacts, etc.

*2.1.3.3. Granular Media close to the Jamming transition*

Common experience indicates that as the volume fraction of hard grains is increased beyond a certain point, the system jams, stops flowing and is able to support mechanical stresses. The dynamical behaviour of granular media close to the 'jamming transition' is very similar to that of liquids close to the glass transition. Indeed, granular media close to jamming display a similar dramatic slowing-down of the dynamics as well as other glassy features like aging and memory effect. One of the main features of the dynamics in glass-forming systems is what is usually called the cage effect, which accounts for the different relaxation mechanisms: at short times, any given particle is trapped in a confined area by its neighbours, which form the so-called effective cage, leading to a slow dynamics; at sufficiently long times, the particle manages to leave its cage, so that it is able to diffuse through the sample by successive cage changes, resulting in a faster relaxation. Contrary to standard critical slowing down, this slow glassy dynamics does not seem related to a growing static local order. For glass-formers it has been proposed instead that the relaxation becomes strongly heterogeneous and dynamic correlations build up when approaching the glass transition. The existence of such a growing dynamic correlation length is very important in revealing some kind of criticality associated with the glass transition.

One can, for example, study the dynamics of a bi-disperse monolayer of disks under two different mechanical forcings, i.e. cyclic shear and horizontal vibrations. In the first case, a "microscopic" confirmation of the above similarity has been obtained and the second can provide the experimental evidence of a simultaneous divergence of length and time scales precisely at the volume fraction for which the system loses rigidity (jamming transition).

## 2.1.4. Bifurcations in turbulence: from dynamo action to slow dynamics

*2.1.4.1. Dynamo action*

Dynamo action consists in the emergence of a magnetic field through the motion of an electrically conducting fluid. It is believed to be at the origin of the magnetic fields of planets and most astrophysical objects. One of the most striking features of the Earth's dynamo, revealed by paleomagnetic studies, is the observation of irregular reversals of the polarity of its dipole field. A lot of work has been devoted to this problem, both theoretically and numerically, but the range of parameters relevant for natural objects are out of reach of numerical simulations for a long time to come, in particular because of turbulence. In industrial dynamos, the path of the electrical currents and the geometry of the (solid) rotors are completely prescribed. As this cannot be the case for the interiors of planets or stars, experiments aimed at studying dynamos in the laboratory have evolved towards relaxing these constraints. The experiments in Riga and Karlsruhe showed in 2000 that fluid dynamos could be generated by organizing favourable sodium flows, but the dynamo fields had simple time dynamics. The search for more complex dynamics, such as exhibited by natural objects, has motivated most teams working on the dynamo problem to design experiments with less constrained flows and a higher level of turbulence. In 2006, the von Karman sodium experiment (VKS) was the first to show regimes where a statistically stationary dynamo self-generates in a fully turbulent flow. It then evidenced other dynamical regimes for the first time, including irregular reversals as in the Earth and periodic oscillations as in the Sun.

These complex regimes, involving a strong coupling between hydrodynamic and MHD, need to be studied in detail. In particular, they reveal that although the dynamo magnetic field is generated by the turbulent fluctuations, it behaves as a dynamical system with a few degrees of freedom.

Theoretical predictions regarding the influence of turbulence on the mean-flow dynamo threshold are scarce. Small velocity fluctuations produce little impact on the dynamo threshold. Predictions for arbitrary fluctuation amplitudes can be reached by considering the turbulent dynamo as an instability (driven by the mean flow) in the presence of a multiplicative noise (turbulent fluctuations). In this context, fluctuations can favour or impede the magnetic field growth, depending on their intensity or correlation time. We can use direct and stochastic numerical simulations of the MHD equations to explore the influence of turbulence on the dynamo threshold.

*2.1.4.2. Bifurcations in turbulence*

At high Reynolds numbers, some systems undergo a turbulent bifurcation between different mean topologies. Moreover, this turbulent bifurcation can conserve memory of the system history. These aspects of the turbulent bifurcation recall classical properties of bifurcations in low-dimensional systems, but the bifurcation dynamics is really different, probably because of the presence of very large turbulent fluctuations. Future studies will be concerned with the universal relevance of the concept of multistability in average for states of highly fluctuating systems and by the transitions between these states (e.g. magnetic inversions of the Earth, climate changes between glacial and interglacial cycles). The slow dynamics of turbulent systems, in the case where exchanges of stability can be observed for some global quantities or some averaged properties of the flow, should also be studied, and an attempt made to construct nonlinear or stochastic models of those transitions.

In the case of turbulent flows with symmetry, it is also possible to construct a statistical mechanics, and to develop a thermodynamic approach to the equilibrium states of axisymmetric flows at some fixed coarse-grained scale. This allows the definition of a mixing entropy and derivation of Gibbs states of the problem by a procedure of maximization of the mixing entropy under constraints of conservation of the global quantities. From the Gibbs state, one can define general identities defining the equilibrium states, as well as relations between the equilibrium states and their fluctuations. This thermodynamics should be tested in turbulent flows, e.g. von Karman flow. Effective temperatures can be measured and preliminary results show that they depend on the considered variable, as in other out-of-equilibrium systems (glass). Finally, we can derive a parameterisation of inviscid mixing to describe the dynamics of the system at the coarse-grained scale. The corresponding equations have been numerically implemented and can be used as a new subgrid scale model of turbulence.

# 2.2. From molecules to organisms

**Reporter:** Christophe Lavelle (IHES)

**Contributors:** Pierre Baudot (ISC-PIF), Hugues Berry (INRIA, Saclay), Guillaume Beslon (IXXI-LIRIS, Lyon), Yves Burnod (INSERM, Paris), Jean-Louis Giavitto (IBISC, Evry), Francesco Ginelli (CEA, Saclay), Zoi Kapoula (CNRS, Paris), Christophe Lavelle (Institut Curie, Paris), André Le Bivic (CNRS SDV, Marseille), Nadine Peyriéras (CNRS, Gif s/ Yvette), Ovidiu Radulescu (IRMAR, Rennes), Adrien Six (UPMC, Paris).

**Keywords:** Systems biology and integrative biology, stability, fluctuation, noise and robustness, physiopathology, biological networks, computational biology, multiscaling in biological systems.

**Introduction**

Biological investigations provide knowledge and are expected, at some point, to translate into clinical research and medical advances for the treatment of human physiopathology. We hope to find cures for diseases and other key medical conditions, if possible, or at least to understand those conditions better. Yet it is increasingly clear that better understanding can only arise from a more holistic or integrative view of biological systems. We thus need to develop a better grasp of biological systems as complex systems, and to transfer this understanding into clinical research. Doing so requires a strongly interdisciplinary approach, and should provide novel insights into physiology and pathology.

After a brief presentation of the general aims and concepts discussed in this topic, we list and offer details on four main challenges. How investigations should be driven in biology is a matter of debate. Should they be data-driven, object-driven or hypothesis-driven? Do we at least agree about the aim of deciphering the causal chains underlying biological processes? Do we expect models to bring insights and knowledge about the behaviour of biological systems, and to make accurate predictions?

Recent advances in functional genomics and in the study of complex diseases (such as cancer, autoimmunity or infectious diseases, mitochondrial diseases or metabolic syndrome) have shown the necessity for an alternative way of thinking in biology, a view in which pathology and physiology result from interactions between many processes at different scales. The new scientific field of systems biology has emerged from this perspective; it focuses on the study of gene, protein, and biochemical reaction networks and cell population dynamics, considered as dynamical systems. It explores the biological properties resulting from the interaction of many components, investigating processes at different scales and their overall systemic integration. Complex systems science provides a conceptual framework and effective tools for unravelling emergent and immergent features from molecules to organisms and vice versa. The term "immergence" is meant to imply that some macro-level constraints cascade back in a causal way onto micro-levels. Both emergent and immergent properties should be understood from the multiscale reconstruction of data recorded at the appropriate spatial and temporal scales. We expect to find generic processes (design patterns for computer science) which apply from upper to lower levels of organization, and vice versa, and which allow their coupling e.g. synchronisation, reinforcement, amplification, inhibition, achieved through basic processes such as signalling through molecular interactions, diffusion, vesicular transport, ionic transport, electric coupling, biomechanical coupling and regulation of molecules and macromolecules characteristic features (including their concentrations).

Complex systems almost always involve a wide range of scales both in time (typically femtoseconds in chemical reactions, seconds in metabolism processes, days to months in cells, and years in an living organism) and space (typically nanometers for molecular structures, micrometers for supramolecular assemblies, organelles and cells, centimeters for tissues and organs, and meters for organisms). Finding the pertinent space and time scales for experimentation and modeling is a major issue. Classical approaches (biochemistry, cellular and molecular biology, behavioural and cognitive studies, etc.) usually have a "preferred" scale set by default, mainly due to the principle protocols and experiments being designed to work only at a specific scale. This makes back and forth interactions between different scales in observations, experimentations, models and simulations a very exciting transdisciplinary challenge.

Variation in biological systems raises the issue of an average, typical or representative behaviour. Determining such quantities, and knowing if they are scientifically useful, requires characterizing and measuring variability and fluctuations at the molecular, single cell, cell population and physiological levels. The origin and functional significance of fluctuations in biological systems, even the scales of space and time on which they occur, remain largely unknown. Their functional significance might be approached through their multiscale transmission and possible amplification, reduction/damping or role in mediating bifurcations.

Obviously, understanding will not arise from a one-to-one description and modeling of organisms (virtual cell, virtual organism) but rather from the correct identification of which components are relevant for a given problem and the reconstruction of models focused on the mechanisms involved. Such a reconstruction should use mathematical and physical tools, some borrowed from out-of-equilibrium thermodynamics and dynamical systems. New tools will also be required to answer specific questions of biology. Ultimately, injecting systemic vision and using complex systems principles and conceptual frameworks for a better understanding of human physio-pathology could lead to novel differential diagnosis and improve medical care.

**Main Challenges**
1. Fluctuations and noise in biological systems
2. Stability in biology
3. Multiscaling
4. Human physiopathology

## 2.2.1. Fluctuations and noise in biological systems

Modern biology has in its development depended heavily on the notion of average behaviours and average individuals. But this conceptual framework has recently been challenged by empirical observation. Quantitative measurements of living single cells, or within such cells, have revealed extensive variability and fluctuation of cellular dynamics between different cells or between different times within the same cell. These observations open a new conceptual framework in biology, in which noise must be fully considered if we are to understand biological systems; this view departs from the classical framework which considered noise and fluctuations it mere measurement error or as "simple" thermodynamic fluctuations which should be suppressed by cells.

This new point of view raises many questions, as well as both practical and theoretical issues likely to deeply modify our understanding of biological systems. However, to tackle these questions, we need to develop a complete scientific program of investigation ranging widely from precise measurements through to analysis of the origin and functional role of stochasticity in biological systems. Among the main breakthroughs, we need to:
- Improve the technology for quantitative measurements of noise and fluctuations in single cells, cell populations, tissues, organs and individuals. In particular, it will be necessary to identify the characteristic times at each level of organization and the most appropriate experimental indicators.
- Identify the mechanisms by which noise and fluctuations arise in biological systems. In particular, what are the modalities of multiscale transmission of fluctuations? Are fluctuations amplified or reduced/damped from one scale to the others? Are they important with respect to bifurcations in the organism/cell fate?
- Understand the functional significance of fluctuations in different biological systems. For instance, it has been proposed that fluctuations can enhance the robustness of living beings. However, other processes can be envisaged (e.g. stochastic resonance, increased signaling rates, cell differentiation, evolution, etc.). Such a functional significance supposes that biological systems are able to control the level of noise.
- Delineate possible mechanisms by which biological systems may control their level of fluctuation (negative/positive feedback loops in biochemical networks, neuronal adaptation in cortical networks, adaptive mutations and mutation hotspots, regulations and networks in the immune system).
- Question the meaning of usual averaging processes in experimental biology. In the case of biochemical networks, can data gathered on cell populations be used to infer the actual network in a given single cell? Similar issues arise in the case of connectivity structures of cortical networks and cell lineage reconstruction.

These issues can be addressed in various biological systems including (but not limited to):
- Transcription and regulation networks: it is now clear that the transcriptional activity of the cell is highly stochastic. Some of the molecular causes of this stochasticity have been identified, yet its precise origin and regulatory mechanisms remain to be discovered. Doing so will first require the development of adequate measurement methodologies to enable us to quantify these fluctuations at different time scales in single cells.
- Neurons and neuronal networks: the so-called "on-going" activity within cortical circuits is a spontaneous activity generated by the recurrent nature of these networks. It has long been considered a mere noise added to the environmental signals. However, more recent studies have proposed a real functional role in which ongoing activity could facilitate signal spreading and be implicated in adaptive processes. Inhibitory effects have been shown to reduce variability at both the single-cell and population level.
- Diversity of the immune system: The immune system is characterized by diversity at different levels. Lymphocyte receptor diversity, populations of effectors and regulators, cell-population dynamics, cell selection and competition, and migration through the whole organism are the result of stochastic or selection mechanisms whose impact on the overall efficiency of the system needs to be further characterized.
- Uncontrolled variability is often accused of being a source of major perturbations in the fate of organisms. Examples can be found in the process of aging, cancer, autoimmunity, infections or degenerative diseases. Yet the precise influence of noise is still open to debate. In particular, one point is to determine to what extent degenerative processes are a consequence of noise accumulation, a variation in noise properties or of rare stochastic events.
- Variability at the genetic level is the major engine of evolution. But genetic variability may be indirectly regulated according to the spatio-temporal characteristics of the environment (selection for robustness, for example, or for evolvability). Moreover, clonal individuals may be very different from each other due to intrinsic and extrinsic phenotypic variability. The mechanisms by which heritable and non-heritable variability are regulated still need to be characterized and their influence on the evolutionary process is largely unknown.

Concerning the modeling of fluctuations, several mathematical and physical tools exist, but these need to be improved. Thus:
- **Stochastic models are largely used in molecular systems biology.** The simulation algorithms (Gillespie algorithm) use the Delbrück-Bartholomay-Rényi representation of biochemical kinetics as jump Markov processes. In order to improve the performance of these methods (which are costly in time) several approximate schemes have been proposed, for instance the approximation of Poisson variables by Gaussians (tau-leap method). Hybrid approximations are more appropriate when the processes are multiscale and these approximations could be developed by combining averaging and the law of large numbers. In certain simple cases, the master equation can be exactly solved.
- **It is also interesting to transfer ideas from statistical physics to biology**. For instance, fluctuation theorems, which concern the occurrence of out-of-equilibrium fluctuations in heat exchanges with the surrounding environment and work theorems, concerning thermodynamic fluctuations in small systems close to equilibrium, could be applied to characterize fluctuations in gene networks, DNA transcription processes and the unfolding of biomolecules.

### 2.2.2. Stability in biology

We encounter various definitions of stability depending on the phenomenon, the model or the community proposing the concept. Frequently invoked concepts include homeostasis in relation to metabolic control, the Red Queen concept in evolution describing continuous development to sustain stable fitness in a changing environment, robustness in systems biology referring to insensitivity with respect to perturbations, or canalization and attractors in developmental biology and ecology.

The main challenges are:

1) In seeking to understand the stability of biological systems, which are always subject to both intrinsic and extrinsic perturbations, we need to develop the notion of steady state, or more generally attractor. We need new mathematical concepts to capture the subtleties of biological stability.

- Finite-time stability is a concept that can be used to define stability in the case when the system is known to operate or to preserve its structure unchanged over a finite time. We are interested in the conditions under which the system's variables remain within finite bounds. Can we extend such formalism to other properties (oscillations, optimal biomass production, etc.)?
- Finite time stability depends on the existence of subsystems with different relaxation times. It is thus important to develop methods allowing to estimate the largest relaxation time of subsystems. For compound systems, how can we relate the relaxation times of the elements to that of the system?
- The notion of resilience is also a generalization of stability that is particularly appealing in this context. Indeed, it focuses on the ability to restore or maintain important functions when submitted to perturbations. The formalizations of this concept, founded on dynamical system properties (measure of attraction basin sizes), or even on viability theory (cost to return into a viability kernel) should become more operational to favour a wider diffusion.

2) The functioning of multicellular organisms occurs at the level of the cellular population, not of the individual cell. Furthermore, the stability of a cell population (tissue) is generally different from that of the individual cell. Cells extracted from tumours, for example, can reverse to normal activity when injected into healthy tissue. In this context, how can we define and study the stability of a population in relation to the stability of individuals? In addition, the same relation should be considered in the context of a developing organism taking into account differentiation and organogenesis. These processes are examples of symmetry-breaking, and we would like to determine whether symmetry arguments can be used in the study of stability properties.

3) Systems biology studies robustness as an important organizing principle of biological systems. As pointed out by H. Kitano, cancer is a robust system with some points of fragility. Thus, finding treatments and cures for diseases may consist in determining the fragility points of a robust system. In order to answer this question, we need good models, new mathematical theories and computer tools to analyse properties of models and new experimental techniques to quantify robustness.

4) Complexity and stability. In the modeling process, we should be able to zoom in and out between various levels of complexity. Stable properties of the system could be those that are common to several levels of complexity. More generally, is there a connexion between stability and complexity?

### 2.2.3. Multiscaling

Biological processes involve events and processes taking place over many different scales of time and space. A hierarchical relationship among these scales enters our description only because it corresponds to our subjective views, usually based on our limited experimental access to the system. Multiscale approaches drawn from theoretical physics have been developed essentially in an unidirectional (bottom-up) way, to integrate parameters and mechanisms at a given scale into effective, and hopefully reduced, descriptions at higher scales. However, lower-scale properties are directly coupled with properties of the higher scales (e.g. 3D chromosome distribution in the nucleus partly governs gene expression, which itself participates in nuclear architecture). The very complexity of living systems and biological functions lies partly in the presence of these bidirectional feedbacks between higher and lower scales which have become established during the course of evolution. Self-consistent or iterative "up-and-down" approaches therefore need to be introduced to account for the strong interconnections between the levels and ensuing circular causal schemes.

*2.2.3.1. Multiscaling vs. self-scaling*

To properly account for the behaviour of a biological system, a multiscale approach should jointly tackle all the scales, with no way to skip any microscopic details or macroscopic assemblies. Obviously, such modeling would rapidly reach a high level of complexity, and would ultimately be intractable. This limitation on multiscale descriptions imposes a drastic challenge to the paradigm underlying the modeling of biological systems.

To reduce the level of complexity, it has been proposed (Lavelle/Benecke/Lesne) to devise models taking the biological function as a starting point and guideline for directing integrated modeling and using supervised data analysis to parallel the biological logic. Decomposition is achieved by dissecting its logic and implementation into basic processes involving features at different scales and are already integrated in their formulation. More generally, such a decomposition results in "self-scaled" functional modules, independent of the arbitrary description or observation scale. As function-dependent representations are inherently multiscale in nature, and the function cannot be discontinuous, this paradigm-transition consequently requires a scale-continuous model. Scale-continuous descriptions may at first sight look prohibitively complex and non-realistic; however, when such a scale-continuous model is constructed in the context of a function-dependent representation, the dimensionality of the variable-vector to be considered collapses.

*2.2.3.2. Emergence vs. immergence*

Modeling of biological systems requires new mathematical formalisms capable of reflecting the complete dynamics of a system by integrating its many levels. This can be achieved by defining "micro to macro" (emergence) and "macro to micro" (immergence, microemergence or downward causation) functions and integrating intra-level (horizontal) and inter-level (vertical) couplings. The definition of pertinent variables at each level of organization, and a description of their relations, is necessary to obtain emergence (resp. immergence) functions that allow analysis to jump from a microscopic (resp. macroscopic) level to a macroscopic (resp. microscopic) level. Emergence and immergence phenomena are well-known in biology, such as the links between the structure topology of tissues and cell behaviour. But these causal relationships are difficult to decipher, mainly because the scales at which they occur are not necessarily those at which researchers make observations and do experiments.
- How should we select relevant space and time scales in our experiments/models/theories (self scaling rather than exhaustive multiscaling)?
- How can we perform multiscale reconstructions from data recorded at different scales? On which spatial and temporal scales will the model/simulation obtained be valid?

## 2.2.4. Human physiopathology and animal models

Human physio-pathology creates uncertainties with constantly moving frontiers between disciplinary fields such as neurology, neurosciences, psychiatry, immunology, cardiovascular, metabolism and endocrinology. Human patho-physiology is characterised by the progressive dysfunction and deterioration of processes acting on multiple space and time scales with non-linear interactions between physiological/biological functions, cognition, emotions and social consequences. Problems can result initially from local conflict between internal and external signals (e.g. dizziness), but this conflict can expand, diffuse and create additional causal loops with multiple pathogenic reciprocal interactions. Functional problems can be primary or secondary effects of spontaneous adaptive mechanisms aiming to counter primary injury and dysfunction, and it is important to dissociate them.

Two main challenges are:
- To apply complex systems principles and theoretical frameworks to the design of experimental studies and the analysis of data at different scales (neurological, physiological, behavioral, neuro-psychological, immunological) from individual or large patient populations;
- To search for cross-correlations and interactions in order to obtain new insights into pathogenic primary or secondary mechanisms. This could lead to new, more sensitive differential diagnostic tools, but also to better medical care or functional re-adaptation. There is a need to go beyond a limited multi-disciplinarity of parallel different approaches and use complex systems tools to combine data from different fields and gain further insight.

This issue concerns the whole of internal and general medicine, immunology, neuroscience, psychiatry, geriatrics, pediatrics, functional re-education, public health, and complex systems science. Examples of functional problems, some of which have no measurable organic basis, include vertigo - dizziness and equilibrium problems and fear of falling in the elderly, isolated hearing loss, tinnitus, learning problems – dyslexia, and also neuro-degenerative diseases, types of dementia, Lewy-Body and Alzheimer's diseases. What causes the switch from physiological auditory noise to perceived unwanted signal in the case of tinnitus in the absence of neuro-ontological findings?

Major questions include the significance of instantaneous fluctuations of measurements (physiologic, behavioural, e.g. in the case of dementia) in relation to patho-physiology and progressive degeneration of cortical-subcortical circuits. Other examples could be given in immunology: analysis of the functionalities of the immune system in physiological (ontogeny to aging, gestation) and pathological conditions (cancer, autoimmunity, infections), and interactions with other biological systems such as the nervous, endocrine, metabolic systems. This is based on dynamical analysis of fluid lymphoid cell populations, quantification and identification of phenotype and functions, repertoires, genomics and proteomics.

# 2.3. Physiological functions

**Contributors:** Catherine Christophe (INRA), Christophe Lecerf (Ecole des Mines Ales), Nadine Peyrieras (Institut de Neurobiologie CNRS), Jean Sallantin (CNRS LIRMM)

**Keywords**: in vivo observation and measurement devices, spatial and temporal multiscale observations, subcellular and supra-cellular functions, organism-environment interaction, ontogenesis, physiological disorders.

**Introduction**

Physiological functions result from the integration of cells, tissues and organ properties in the context of the whole organism interacting with its environment. A complex system approach to physiological functions should lead to an iterated cycle combining relevant measurements and experimentation, modeling and simulation. Such a goal requires building multimodal investigation devices for simultaneous *in vivo* recording at different spatial and temporal scales of relevant parameters as well as designing theoretical methods and tools for appropriate modeling and computer simulation.

Expected results include the design of new investigation devices and theoretical methods and tools for observing, modeling, understanding and possibly controlling physiological functions.

> **Main Challenges**
> 1. Integrating multimodal measurements and observations of physiological activities at different spatial and temporal scales.
> 2. Characterizing the contextual features which determine the onset of a physiological function, or their maintenance and modulation.
> 3. Investigating the relationship between the ontogenesis of a physiological function and its potential disorders.

## 2.3.1. Integrating multimodal measurements and observations of physiological activities at different spatial and temporal scales.

An integrated observation of sub cellular and supra cellular processes requires either:
(i) To translate in the same spatial and temporal reference frame heterogenous data recorded in the same organism but at different moments, or
(ii) To design new devices capable of simultaneously recording multimodal data.

The first goal can be achieved through available methods going from spatio-temporal matching to data fusion. These methods are limited by recalibration problems and errors (whatever the rigid or elastic transformations applied).

The second option would be a real breakthrough and provide a generation of totally new instrumentation offering instantaneous access to essential structural and dynamic variables (chemical, electrical, mechanical, etc.) at all relevant spatio temporal scales. The trend in this direction is exemplified by macroscopic data acquisition in medical imaging with optical-PET and PET-CT devices and, for vital physiological variables, by ambulatory integrated sensors providing real-time patient state tracking in a normal environment. In the domain of vegetal biology, phenotypic plant platforms lead to the observation of flow from roots to leaves at different time scales. Integrating such synchronous, multimodal, multiscale observations in relevant models should provide a good basis for the reconstruction of physiological functions.

## 2.3.2. Characterizing the contextual features determining the onset of a physiological function, or its maintenance and modulation.

The objective is here to view the function as an integration of subfunctions which can be investigated from different perspectives or using perturbative and comparative approaches. Different factors or conditions, such as resting versus moving, diet-nutrition, or training, can influence and move the system towards new functioning modes. Comparative physiology provides a way to study the conservation or divergence of physiological functions. This approach is relevant for respiration and locomotion in the animal kingdom as well as for fruit maturation in the field of vegetal biology.

Physiological functions should be characterized through the extraction of high-level variables, loosely akin to "thermodynamics variables", or along the lines of allometry i.e. preservation of characteristics over the size variations. More generally, we should be able to define invariants (or invariant relationships) attached to physiological functions and the conditions for their conservation.

## 2.3.3. Investigating the relationship between the ontogenesis of a physiological function and its potential disorders.

Physiological functions should be explored through their set up during ontogenesis, maturation and maintenance during growth, adulthood and ageing. The dynamical behaviour of physiological functions should be explored also during pathological events.

**Examples:**

- **Heart embryology:** progressive formation of anatomical structures and functional patterns with ill-posed problems related to the partial observations at our disposal (i.e interpolation of objects with high structural variation from the architectonic viewpoint, installation of nodal tissue functions or sinusal electric waves, etc.)
- **Schizophrenia:** effects on the highest cognitive levels of the modifications induced by the disease at the level of more elementary neurological functions

# 2.4. Ecosystemic complexity

**Contributors:** Olivier Barreteau (Cemagref), Paul Bourgine (Ecole Polytechnique), David Chavalarias Institut des Systèmes Complexes de Paris Ile-de-France / CNRS), Cédric Gaucherel (INRA), François Houllier (INRA), Ioan Negrutiu (ENS Lyon), Nadine Peyrieras (Institut de Neurobiologie CNRS).

**Keywords:** ecological dynamics, adaptation and evolution, ecological services, multi-functionality of the ecosystems, integration of data, coupling of models, space-time dynamics, multiscale models, disturbance and resilience, stability and dynamic transition, emerging behaviour, feedback and retroaction, functional organization.

**Introduction**

Defined as the close association of an abiotic environment and a collection of living organisms, an ecosystem is characterized by a great number of physicochemical factors and biological entities which interact with each other. The multiplicity and diversity of these interactions as well as their involvement of a vast range of levels of organization and a broad spectrum of space and temporal scales justify the expression of "ecosystemic complexity."

Moreover, ecosystems, whether natural, managed or artificial, are subjected to "perturbations" (e.g. natural hazards or biotic and abiotic stresses) and deliver many and diversified commercial and non-commercial products and "services." To account for this ecosystemic complexity, to understand the resilience of the ecological processes and to open the possibility of ecosystem management and control, we need to articulate various strategies -- for reconstructing the spatial and temporal dynamics, starting from observations and from increasingly instrumented experiments, for theoretically and experimentally identifying the retroactive mechanisms and emergent phenomena, and for modeling and validating these models.

> **Main Challenges**
> 1. Develop observational and experimental systems for the reconstruction of the long-term dynamics of ecosystems.
> 2. Model the relationships between biodiversity and the function and dynamics of ecosystems.
> 3. Associate integrative biology and ecology to decipher evolutionary mechanisms.
> 4. Simulate virtual landscapes (integration and coupling of biogeochemical and ecological models into dynamic landscape mock-ups).

## 2.4.1. Develop observation and experimental systems for the reconstruction of the long-term dynamics of ecosystems

The rapid development of *in situ* systems of measurement (metrology and sensors) is making possible the integration of data collected within networks of observation (spatial and temporal sampling strategies, environmental research observatories) and/or of experiments (microcosms, mesocosms) in models of ecosystems. Further progress requires the development of information systems based on a conceptual modeling of studied ecosystems and tools for the multidimensional analysis of data coming from multiple sources ("meta-analysis").

## 2.4.2. Model the relationships between biodiversity, functioning and dynamics of the ecosystems

These relations, which play a central part in the vast field of biodiversity studies, describe various functions (production, transfers of matter and energy, resistance and resilience to perturbations, etc.), at different scales of space (station, landscape, area, continent) and of time. Historically, researchers have approached the study of these relations either by wondering about the way the environment and the functioning of the living organisms and their interactions determine the assemblies of species, or, more recently, and in a reciprocal way, by studying the role of the richness and specific diversity in the way ecosystems function.

## 2.4.3. Associate integrative biology and ecology to decipher evolutionary mechanisms

To understand and model the response of ecological communities (in their structure, functioning and dynamics) to the changes of their environment (climatic changes, pollution, biological invasions, etc.) we require a better comprehension of the adaptive mechanisms. This task can now be supported by conceptual, methodological and technological progress made in integrative biology (genomic functional calculus, biology molecular, genetic, physiology and ecophysiology) and by the convergence of approaches from population, molecular and quantitative genetics.

## 2.4.4. Simulate virtual landscapes (integration and coupling of biogeochemical and ecological models into dynamic landscape mock-ups)

The concept of virtual mock-ups, based on a categorical representation of the landscape mosaic, would make it possible to build a typology of representative landscapes (hedged farmland, open field, mixed landscapes, forests, peri-urban areas, etc…). The following phases would consist in modeling, first, the functioning of the landscape (i.e. biogeochemical cycles, transfers and exchanges: air particulate transport, determinism of the microclimate, transport of water and of associated pollutants in the soil and the watersheds) with as a deliverable the production of functional relations between landscape topology and structure of the exchanges. Second, it is also important to model the dynamics of the landscape (i.e. evolution of its space organization) under the effect of the human activities and of certain ecological processes (for example, colonization of spaces by the vegetation). Such a tool would have a great utility in ecology or epidemiology, in the agronomic disciplines and for the local management of land and land use.

## 2.4.5. Design decision-support systems for multifunctional ecosystems

All decision making linked to the ecosystem management would be greatly assisted by the qualification and quantification of the products and services provided by the ecosystems, and the integration of these services and products in systems of policy-relevant indicators (dashboards, tools of decision-making assistance, life cycle analysis and eco-balance analysis, etc.). Policy formulation and implementation also require much more sophisticated modeling and quantification of human practices and techniques, or management systems relating to ecosystems, and fully coupled models taking into account stochastic components (whether those are intrinsic or that they are related to the incomplete character of knowledge on the elements of these systems, their interactions and the extrinsic factors likely to disturb them).

# 2.5. From individual cognition to social cognition

**Reporter**: David Chavalarias (Institut des Systèmes Complexes de Paris Ile-de-France / CNRS)

**Contributors**: Paul Bourgine (Ecole Polytechnique), David Chavalarias (Institut des Systèmes Complexes de Paris Ile-de-France / CNRS), Jean-Philippe Cointet (Institut des Systèmes Complexes de Paris Ile-de-France / INRA ), Guillaume Deffuant (Meteo-France & U. Paris-Est Cemagref), Camille Roth (CAMS).

**Keywords:** Social dynamics, decision criteria modeling, quantitative social measurement, social cognition, inter-individual heterogeneity.

## Introduction

Cognition is information processing, understood in a wide sense; that is, including all related aspects such as, for instance, interpretation processes. A cognitive system is thus an information processing system. It can be embedded in a single individual or distributed over a large number of individuals. We speak of individual cognition or distributed cognition. Social cognition is a cognitive process distributed over all members of a society, interacting within a social network. Individual cognition as well might be considered as distributed cognition over a neural network.

In social networks, as information reaches agents, its content is then processed by the social network, producing other pieces of information and other social links following series of interactions. This process of social cognition could thus lead to a transformation of the social network.

At the individual and collective levels alike, cognitive processes obey strong constraints: individuals cannot achieve anything outside of what they are able to do themselves or in interaction with others; nothing can be anticipated outside of what they can predict alone or by interacting with others. Both the network structure and the nature of interactions act as strong constraints on cognitive processes. New protocols appear which make it possible to describe or quantify these constraints at the infra-individual, individual and collective levels, thus suggesting, in turn, new models. The quick migration of social interactions towards digital media enables the massive collection of data on social cognition, from the viewpoint of both its processes (spatial structure of interactions, temporal distributions, etc.) and its products (online documents, user-focused data, etc.). The coexistence of these two phenomena opens today new perspectives for the study of individual and social cognition on the basis of benchmarking models with empirical data. This ought to be a major focus of research initiatives for a better understanding of the evolution of our societies.

> **Main Challenges**
> 1. Individual cognition, cognitive constraints and decision processes
> 2. Modeling the dynamics of scientific communities
> 3. Society of the Internet, Internet of the society

## 2.5.1. Individual cognition, cognitive constraints and decision processes

The relationship between high-level and low-level cognitive processes remains an unsettled issue: the link between dynamic processes in the neural network and symbolic processes as they are being studied by psychology and linguistics is still open to question. A promising approach consists in exploring in a much more precise manner meso-scale spatio-temporal dynamics, such as, for example, cortical columns or synchronized neural assemblies (or, more broadly, polysynchronic assemblies). These spatio-temporal dynamics may be useful in elucidating the microscopic dynamics behind symbolic processes. In order to understand better the links between dynamic and symbolic processes, further theoretical and methodological exploration, as well as sharing data from very large databases provided with their metadata, is required.

Significant progress towards this challenge would not only lead to unifying an essential aspect of cognitive science, but would also launch much more strongly the new discipline of neuroeconomics: observing neural activity brings a novel viewpoint on the study of human behaviour towards "nature" or in relation with strategic and social interactions with other individuals. From the perspective of cognitive economics, this brings hopes that decision theory could be revisited, as well as standard game theory, including the notions of "preference" and "utility" which are foundational for economic theory.

## 2.5.2. Modeling the dynamics of scientific communities

Scientific communities constitute a privileged area for the study of social cognition because both the structure of the underlying networks (team organization, collaboration networks, co-authorship networks, or citation networks) and the production of these communities (conferences, journals, papers) is known in a dynamic fashion. In order to exchange concepts, scientific communities create their own language whose evolution reflects their own activity.

This makes it possible to address very precise topics pertaining to how these scientific communities are collectively processing information. How are new concepts or issues being adopted? What are the dynamics by which innovations diffuse (effect of authorities, local traditions, etc.)? What is the effect of the breakdown of individuals in communities or the creation of links between communities on the development of knowledge? Which are the relationships between individual trajectories and community evolutions? What tools should we create to visualize dynamically the evolution of scientific paradigms, taking into account the continuing input of scientific production?

**Examples:**

- Emergence and diffusion of new concepts in bibliographical databases
- Detection of emerging scientific fields
- Dynamics of collaboration networks
- Paradigmatic comparison of distinct scientific communities or institutions

## 2.5.3. Society of the Internet, Internet of the society

The quantity of information stored on the Internet will have soon easily surpassed that stored on paper. The Internet concentrates today various types of knowledge storage systems (papers, encyclopedias, etc.). It is also a place where discussions (weblogs, forums) and commercial transactions (auction and trade websites) occur, referencing is being produced (for individuals through personal webpages as well as for institutions and organizations), and it serves as an external memory for relationship networks (friendship networks, work groups, etc.). It is also a " world agenda" with hundreds of thousands of events being announced every day. What modifications is this new tool bringing to processes of social cognition (new kinds of encounters, new kinds of exchange, new kinds of debates, new kinds of collective building of knowledge)? For the first time, we may empirically work on this type of data with a fairly large spatio-temporal precision. How can we use these new sources of information to better understand social dynamics and create tools to visualize the complexity of social activity which the Internet is revealing? A major challenge is to transform raw information available from the Internet into structured flows of information which make it possible to visualize, model and rebuild social cognition processes at work on the web.

**Examples:**

- Impact of weblogs in political and civil debates,
- New dynamics for the collective elaboration of knowledge (Wikipedia, open-source software, etc.) ,
- Measuring the propagation of social emotion following important social events, through the number of requests (ex: Google trends) ,
- Comparative study of cultural differences through geo-localized informations (semantics in webpages, tags, requests on search engines, etc.), reconstruction of cultural territories.
- Formation of epistemic communities, friendship networks

# 2.6. Individuals, structures, societies

**Reporter:** Denise Pumain (Université Paris 1)

**Contributors:** Frédéric Amblard (Université de Toulouse), Cyrille Bertelle (LITIS UFR sciences technologies), Paul Bourgine (Ecole Polytechnique), David Chavalarias (Institut des Systèmes Complexes de Paris Ile-de-France/CNRS), Valérie Dagrain (consultante technologie information), Guillaume Deffuant (Meteo-France & U. Paris-Est Cemagref), Silvia De Monte (ENS), Sara Franceschelli (Paris Jussieu), Francesco Ginelli (Institut des Systèmes Complexes de Paris Ile-de-France), Pablo Jensen (ENS Lyon), Maud Loireau (IRD), Jean-Pierre Müller (CIRAD), Denise Pumain (Université Paris 1).

**Keywords**: Institutions, heterarchies, multilevel methodologies, flocking, collective behavior, (evolutionary) game theory, cooperation, quantitative measurements, evolution, perturbation response, spatial organization, social insects, transition to multicellularity, synchronous oscillations, social differentiation, cognitive economics, social networks, social learning.

**Introduction**

Interacting individuals create organisations and are (re)created by them. The behaviour of societies is not the simple sum of its elements, but often displays true emergent properties. For the sake of analysis, we can distinguish between weak and strong emergence. While certain questions – such as the emergence of clustering and/or flocking behavior - can be regarded as common to any biological population, human societies included, culture and reflexivity introduce new challenges to description efforts. Moreover, geographical entities (regions, spatial organisations, networks, landscapes, etc.), influence the interactions between individuals. Complex systems methods provide the theoretical framework to understand the coevolution of the different description levels (territories, societies and individuals) and the evolutionary processes which shape collectives. It helps to analyze the reasons which lead to inequalities between different entities.

This challenging question is especially relevant for understanding the multilevel dynamics of geographical entities: (places, regions, spatial organisations, networks, landscapes, etc.). Most often, stylised facts can be obtained from the co-evolution of territories at the macro-geographical level. Models of complex systems can help to reconstruct individual behaviours which, under given societal rules and historical context, generate inequality between territories. Territorial entities must be understood not only as geographical subdivisions but also as systems with particular governance rules and associated collective representations which define feelings of belonging which characterise individual identities.

Challenges summary: the main research question in this area is to identify which universal properties at the macro level may be explained by collective behaviours (described and quantified from societal surveys). For understanding the process of social (or geographical) differentiation, we require two types of modeling of strong emergence phenomena: firstly of the retroaction of collective patterns on individual representations and practices, and secondly, of the emergence of institutions at collective levels out of the interactions between individuals and changing collective rules. Another very important challenge, when comparing the development of social sciences with the "hard" sciences, and for enabling the transfer of models from studies of physical matters to societal ones, is to properly collect in vivo or in vitro data (adapted statistics, data from experimental economics, etc.)

> **Main Challenges**
> 1. Emergence of collective behavior in biological populations
> 2. Co-evolution of individuals and society
> 3. Co-evolution of individuals, structures and territories
> 4. Heterarchies, multiscale organisations

### 2.6.1. Emergence of collective behavior in biological populations

Animal societies are commonly described at the collective level due to the evidence and immediateness of population-level observations. Often, the characteristics of individual agents shape the collective behaviour in a nontrivial manner. One of the most interesting challenges of the complex systems approach is then to unveil the relationship between the individual and society levels in biological populations.

This is, for instance, the key question when one ponders "flocking" behaviour in animal species such as birds, fish, herding mammals and bacteria. Another fascinating phenomenon is spatial organization which emerges spontaneously in the nesting behaviour and food foraging trails in social insects. At the cellular level, collective synchrony may emerge as a result of the interaction of individually oscillating cells. Moreover, the transition from unicellular to multicellular organizations is a major issue of evolution theory. The study of excitatory (neuron-like) units interacting via nontrivial connectivity graphs has recently shownrich behaviour related to coherence and/or partial synchronization.

In recent years, tools imported from nonlinear physics have supported a considerable effort through theory and modeling to characterize these emergent phenomena. Basic mechanisms leading to collective behaviour have been identified, some universal properties - common to different biological systems - have been documented and new predictions formulated at the theoretical level. While a qualitative agreement with observations has been generally reached, we need quantitative measurements in biological populations to further advance our comprehension of these phenomena.

The emergence of coherence among oscillating cells, the so-called Kuramoto transition, for instance, is expected to occur in a wide class of cellular populations. Yet, it has been quantitatively demonstrated so far only in physical and chemical systems. Superdiffusive behaviour and anomalous number fluctuations have been theorized to occur in flocking groups, but have not be studied in empirical observations. The transition to multicellularity, described by game theoretical models, has never been observed in experiments of directed evolution.

Furthermore, a novel set of questions arises concerning the stability of such emergent social structures with regard to external forcing or perturbations. For instance, has the interaction between birds in a flock been optimized by evolution for resilience against the disruptive effects of an attacking predator? Is it possible to control an entire animal group by influencing a few of its elements? How do ants react to new obstacles introduced into their foraging ground? What is the robustness of synchronous behaviour with respect to individual diversity?

## 2.6.2. Co-evolution of individuals and society

The renewal of transdisciplinary research in the social sciences reflects the need to integrate numerous aspects of human behaviour to fully understand the diversity of human cultures and socio-economic institutions. This is particularly clear in the debate among economists who, after long commitment to the homo economicus paradigm, are increasingly abandoning this view and seeking a new paradigm. Indeed, many phenomena do not fit with traditional explanations of socio-economic equilibria, in particular the observed heterogeneity of socio-cultural patterns, the fact that we often face transitory phases and local attractors instead of stable equilibria and the accumulation of evidence about the importance of social influence and others' beliefs in decision-making processes, even in economic settings. On the other hand, policy-makers are pointing out ever more often the fact that new societal challenges such as global warming or persistent poverty in certain areas require us to address the issue of changing mentalities (i.e. the distribution of preferences or types of agents in population) rather than just changing behaviors. Since mainstream economics, and most of the time formal social sciences, consider preferences as fixed over time, this leads to new theoretical challenges for economics, and for the social sciences more generally, as explained by the Nobel prize-winner Vernon Smith (2005):

*"Technically, the issue can be posed as one of asking how most productively to model agent 'types' by extending game theory so that types are an integral part of its predictive content, rather than merely imported as an ex post technical explanation of experimental results."*

This question about the origin of types, preferences or, in a more comprehensive view, the representations, beliefs and values of agents, is one of the more tricky issues that social system modeling has to tackle. Since all the agents' decision-making processes are derived from them, it is hard to imagine the correct grounding of models and their conclusions without addressing the question. However, very few models tackle the question directly. Moreover, most rely on a mechanism of social conformity to make types of agents evolve, while it is not clear that the diversity of agents' types in a society can be acounted for only by this mechanism.

To go beyond this view, we have to imagine formal frameworks to represent social differentiation where the process of differentiation is neither an optimization of a given quantity nor the sole by-product of social conformity.

We thus have to find an alternative between methodological individualism and holism, where both social influences and individual motivation contribute to the process of differentiation of agent types. A third alternative that could be named *complex methodological individualism* (Jean-Pierre Dupuy 2004).

To make this programme concrete, we need to address several methodological questions:
1) We need to investigate mechanisms driving changes in beliefs, goals, preferences and values from the point of view of psychology, cognitive sciences and philosophy. This will help to develop suitable tools to represent formally the dynamics of change at an individual level. We need to account at the same time for individual agency (the ability to choose according to one's own personality) and for social influence on the evolution of the individual personality. We must bear in mind the fact that people are able to decide not to follow some rules and to create new rules. How can we go from "simplistic" interaction mechanisms (optimum, copy, etc.) towards more creative ones? It may be argued that people, in many important cirumstances, do not (seem to) follow algorithmic rules. Instead, their expertise adapts intuitively to the context: how can we take this into account in simulations?

2) We need to better understand the way people shape their social network, how they form new links, how they prune old ones, how they perceive social groups and to what extent their decisions are socially and spatially embedded. In particular, it is important to succeed in representing social groups as endogenous outcomes of social dynamics rather than as entities given top-down by the modeler, as they are often described in the literature.

3) We need to develop methods to articulate behaviour at the individual level as well as emergent collective behaviours at all scales in space and time. In particular, we should address the question of how emergent behaviour can impact retroactively on individual behaviour (bottom-up and top-down influences) and what influence network topology has on the dynamics a network supports.

4) We need to better understand the roles of history, path-dependence and perturbations. In the real social world, structure is the product not just of contemporary actions but of history. It is there as we act, and our actions both constitute history and change it. In other words, for a simulation to be adequate as a representation of the social sphere, it cannot start from agents alone. It is important to include the history in simulations, and pay attention to path-dependence. Often, this dependence arises from "errors" in the inference and transmission of information, from variations in the environment or from the heterogeneity of individual responses. Studies of dynamical systems reveal that these pertubations are a key component in the determination of system behaviour.

5) Can we express clearly how social reality feeds back retroactively on agents? At the intermediate level of "habits of thought," in Hodgson's terms, "The effect of institutions is to frame, channel, constrain individuals, giving rise to new perceptions and dispositions. From these new habits of thought and behavior emerge new preferences and intentions, changing the institutions, which in turn affect the manner of seeing."

6) What are social models? When evaluating the relevance of a model, it is interesting to estimate the representation of the "social" in the model. The relevance of "simple" models of social systems (in the style of Schelling's segregation model) is an open question which deserves careful epistemological investigation. Possible questions include: to what extent is the reflexivity of the agents taken into account? What intermediary levels between individuals and society (such as institutions, cultures) are taken explicitly into account?

### 2.6.3. Co-evolution of individuals, structures and territories

Territorial entities (towns, landscapes, regions, etc.) are coevolving with individuals and social structures (point 2). Territorial entities act as a context that constrains (both enables and limits) individual capacities, while the individuals and the collective social structures, through their practices and interactions, maintain or transform the territorial structures. The very existence of territories, through the resources they give access to, or the symbols representing them, or the control they have on different aspects of life, acts as a constraint on (as well as a resource in) the evolution of individuals. This coevolution involves processes which take place on different time scales. The relation between space and time scales is not trivial and requires specific investigation.

One problem is to identify the relevant time scale for the observation of spatial entities (for example, in the acquisition of remote sensing data). Another problem to solve is how to identify relevant territorial entities over time and between different territorial systems (for example, identifying a "city" as an agglomeration of urban population).

Another challenge is to identify which processes tend to increase the inequalities between territorial entities (for example, of income or GDP between countries, or population size or total urban GDP between cities), and how positive feedbacks and scaling laws are related to the co-evolution of individuals and social organisations. This raises the issue of the role of the social or cultural or economic diversity associated with large size as compared with the role of specialisation.

How can we understand transition between stages (periods of time, regimes) where the dynamics is constrained by the limitation of local resources (ecological systems), and stages where innovations or the expansion of spatial networks removes constraints on system expansion?

Territorial entities are organised in social and spatial networks through relations depending on the state of communications and transportation. Over historical time, spatial distance has been a heavy constraint on social interactions, even if long-distance relations have always existed. Today, many relationships seem to be no longer or less constrained by distance (and its interpretation as a cost or a length of time). The apparent contraction of space over time seems to increase territorial inequalities. Changes in the configuration of transportation networks have strong effects on the propagation of epidemics, whereas the consequences of changes in communication networks are much more difficult to estimate.

Three main components of the dynamics of territorial entities (such as their growth in demographic and economic terms or their potential capacity in terms of sustainability) are already well-identified, but remain to be quantified as possible factors of sustainable development. These are intrinsic resources (including landscapes, human capital, portfolio of economic activities, value of heritage), geographical situation (relative position in economic, financial, geopolitical or cultural networks, evolving through time), and the path dependence which both enables and limits a subset of dynamic trajectories of individual territories. For instance, it is possible to estimate in probabilistic terms the future of rural localities as depending on their economic specialisation, their geographical situation relative to cities of different size and functions, their own potential resources, and the capacity of initiative of their main decision-makers. In the case of cities and metropolitan areas, the weight of these factors is different. These weights can be estimated from comparative studies of scaling laws for the distributions of city and metropolitan sizes, revealing the incidence of the main innovation cycles on urban development.

A major research challenge in this domain is about finding the right data for quantifying the interactions between territories. Limited data on material interactions between cities, regions and countries (as migration flows or trade exchanges) can be collected, but we often lack data on energetic, financial or information exchanges between them. These invisible flows are actually those which maintain and build the dynamics of unequal growth and development among territories. They reflect and create the interactions generating material flows, yet are generally analysed in a very indirect way, as "network effects", or summarised under "generalised distance measurements" which seek to capture the relative situation of each place within a larger system. Any way of improving knowledge about these crucial energetic, financial or informational interactions, at any level, would enable significant progress in the analysis of territorial dynamics.

Experimental models of urban sustainability could help in developing policies to face the challenges of either an increasing scarcity of energy or a change in the organisation of the global financial and information networks connecting the metropolitan areas around the world. A major question concerns the threat to global stability that will soon emerge from the growing divergence between demographic and economic growth: over the next two or three decades, the largest metropolises in the world in terms of population will also be the poorest in terms of economic production. we need to simulate the reorganisation of global flows at the scale of national territories and of metropolitan networks to estimate the probable costs of sustainability.

## 2.6.4. Heterarchies, multiscale organisations

A first distinction can be made between embedded hierarchies in which the macro-level entities recursively embed the lower level entities, such as hierarchies in ecology (e.g. cells, organs, organisms, etc.), and non-embedded hierarchies where entities are representative of and/or command the lower level entities (as in structures of government). In both cases, the hierarchy depends on the point of view: typically, embedded hierarchies can be observed by looking at the relative time scales of the dynamics, and a higher level entity in a non-embedded system can be a lower level entity in another organization (e.g. representative or not representative). Many questions about complex systems require the combining of heterogeneous embedded and non-embedded hierarchies. A number of abstractions are proposed for dealing with the description of such systems, such as multi-hyper-networks (Jeffrey Johnson) and holonic structures (Koestler).

The remaining questions are:
- how to represent holonic structures and their dynamics;
- how to reconstruct both holonic structures and multi-hyper-networks dynamics from observed data;
- how to represent these structures and their dynamics to perform predictions.

When holons are used, the question is how to manage their Janus's double face (autonomy with regard to the lower level / heteronomy with regard to the upper level) and how to choose the best-suited model of autonomy.

Applications might include accounting for different kinds of hierarchies in modern and traditional cultures, the intertwinning of governance mechanisms on a given territory, etc.

The processes which generate and maintain heterarchical structures (partly embedded and partly intertwined interaction networks) are generally the same. The structure is defined both by qualitative differences and quantitative inequalities between the territorial entities, which are sufficiently pervasive over periods of time longer than the behavioural regime or even the generational renewal of their components. These structures are transformed either smoothly or more rapidly through innovation processes which use part of the structure (and the potential "comparative advantage" of the territory) for introducing new social practices or new artefacts.

To establish better predictions of the territorial capacity to evolve and capture innovation, we need better knowledge of the heterarchical structures. Which methods and instruments can be used for describing and comparing heterarchical organisations, including the number of levels, their degree of flexibility or embeddedness, and their functional articulation?

To be able to predict the capability of adaptation and innovation of a territorial entity, we need an analysis and classification of the historical trajectories of territorial entities, including their sensitivity to internal organizational features and to external perturbations. This means large explorations (data mining and dynamic modeling) of the evolutions of territorial entities in socio-economic and geographical context.

# 2.7. Innovation, learning and co-evolution

**Reporter:** Denise Pumain (Université Paris 1)

**Contributors:** David Chavalarias (Institut des Systèmes Complexes de Paris Ile-de-France/CNRS), Nadine Peyrieras (Institut de Neurobiologie CNRS), Denise Pumain (Université Paris 1).

**Keywords:** innovation, emergence, bifurcation, co-evolution, learning, acceptance, society of information.

## Introduction

Novelty in complex systems appears through a variety of processes including the emergence of new entities and new categories, the modification of interaction processes, changes of their temporal or spatial scales, or their dynamical transformation. Within the perspective of complex systems science, the main question is whether the modes of change are comparable when moving from natural or artificial systems towards social systems. A first challenge is to identify which dynamic conditions are favourable to innovation. Is innovation always associated with jumps, ruptures or bifurcations, or can it proceed from more regular trends? Which processes explain the frequent observation of innovation cycles? A second challenge is to determine whether there is an acceleration of innovation in human society through time, by identifying relevant measures of societal changes. A third challenge is to understand how intention and reflection frame innovation in social systems and how the feedback effect of learning affects individual and collective cognition over historical time.

> **Main Challenges**
> 1. Understanding dynamic conditions of innovation
> 2. Modeling innovations and their rhythms
> 3. Understanding the relation between cognition and innovation

## 2.7. 1. Understanding the dynamic conditions of innovation

Can innovation only be analysed *ex-post*, or can it be predicted, and if so, from which indicators and explanatory variables? Are the signs that announce coming change evident in a specific part of the system's dynamics, through the amplification of fluctuations around a trajectory, intensification of pre-existing processes, or the transition between quantitative toward qualitative variations? How does innovation become accepted, either by introducing itself into existing structures or replacing them, or by inducing modifications of these structures? Which relationships are established between new artefacts and their functionalities, and the new practices based around their use? How can we explain how groups of many innovations lead to the observation of large cycles in social evolution?

## 2.7.2. Modeling innovations and their rhythms

Certain analysts suggest that there is an acceleration of the production frequency of innovations, especially through the technical revolutions and the evolution towards a society of information. Is this reality or illusion? Answering that question requires a rigorous definition of innovation and of information and careful determination of the time intervals that measure its frequency. How can we build reference times that are relevant for characterizing the rhythms of emergence, succession and co-presence of innovations? In other words, is the regular hour time meaningful or should one imagine other measures of societal time?

## 2.7.3. Understanding the relation between cognition and innovation

Societies also build and assimilate innovations concerning the artefacts they produce in their own practices and institutions. Is it possible to understand the social dynamics of innovation without introducing the individual and collective intentionality and reflexivity? Is social innovation in cooperation or conflict with biological evolution? Does the fact that innovation is targeted, and that the processes of learning and acceptance are conveyed through legal, economic or cultural regulations, introduce different characteristics for innovation in human societies? Within these processes, is it possible to identify at meso-levels social milieux or networks or geographical spaces that would be more favourable to innovation, or loaded with a specific innovative capacity? What are the expressions of the interactions between innovation and individual cognition? Can the social control on innovation reach as far as the biological transformations?

# 2.8. Territorial intelligence and sustainable development

**Reporter**: Denise Pumain (Université Paris 1)

**Contributors**: Pierre Auger (ISC-IRD-Geodes), Olivier Barreteau (Cemagref), Jean-Bernard Baillon (Université Paris 1), Rémy Bouche (INRA), Danièle Bourcier (CNRS), Paul Bourgine (Ecole Polytechnique), Elisabeth Dubois-Violette (CNRS), Jean-Pierre Gaudin (IEP), Elisabeth Giaccobino (CNRS), Bernard Hubert (INRA), Jean-Pierre Leca (Université Paris 1), Jean-Pierre Muller (CIRAD), Ioan Negrutiu (ENS Lyon), Denise Pumain (Université Paris 1).

**Keywords:** geographical space, territorial configuration, rural and urban regions, networks, systems of cities, multi-level and multi-actor governance, resources, regulation, sustainable development, negotiation, geographical information systems, cellular automata, spatial simulation, multi-agents systems.

## Introduction

A physical territory is a system that naturally integrates a variety of processes usually analysed by a diversity of disciplines (economics, sociology, and so on). These processes require natural and social resources and include individual and collective actions, which together act in building the territory. Households, firms or government bodies take both planned and unplanned actions, as well as reiterated practices and strategic anticipations. Physical infrastructures as well as immaterial long lasting socio-spatial configurations constrain these actions and also shape the territory at several scales in space and time. Penetrating this complexity requires simulation models -- for understanding the relationship between processes and structures, for evaluating and preparing individual and collective actions, or for measuring their impact on the viability of spatial structures. Such models are important tools for intelligent decision-making and may then contribute to change the evolution of territories.

> **Main Challenges**
> 1. Understanding territorial differentiation.
> 2. Towards a reflexive territorial governance
> 3. Viability and observation of territories

### 2.8.1. Understanding territorial differentiation

Territories are reorganized at different scales, from local to global, through the expansion of material and immaterial networks and the diversification of levels where decisions take place. "Network territories" are today forming through telecommunications-based links, ignoring the need for physical proximity, at the level of individuals and of global firms. At the same time, contiguous territories are partially intersecting, for instance when their future is governed by several decision centres. Are the classical territorial models still valid for representing geographical differences? How can they be replaced?

The evolution of territories is usually described in terms of geo-history, territorial viability, or adaptation and innovation capacity. It must be related to processes such as the development of institutions, technological innovations, transformations of social practices and representations. Within this context, modes of circulation and concentration of information are essential. Very often, networks conveying important information are not observable, and have to be reconstructed through simulation models. The challenge is to couple dynamic models representing spatial interactions at a variety of scales and geographical information systems which can integrate and visualize the located information and the evolution of networks and territories.

## 2.8.2. Towards a reflexive territorial governance

Territorial governance no longer takes place through simple hierarchical top-down control, but a multi-level process involving many actors. Intermediate control structures are emerging between territorial scales. New models of legitimating power are being invented between representative and participative democracy and inclusive governance. Moreover, the growing interest for sustainability invites us to seriously take note of the natural dynamics that operate at different scales of time and space as well.

The building of well-informed governance relies on the invention of new decision models which consider processes and institutions, configurations of competition and cooperation, and also symbolic and practical interactions. Natural and social dynamics have to be coupled in identifying organization levels, scales of time and relevant territorial subdivisions for a reflexive control. A further difficulty is to include the diversity of the strategies of the actors in such models. Generally speaking, the question is to identify which structures are emerging at the meso level and to understand what are the linkages between micro, macro and meso levels.

## 2.8.3. Viability and observation of territories

The retrospective and prospective analysis of territories is essential for improving knowledge about the long-term sustainability of geographical entities in their social, economic, ecological and ethical dimensions. Questions of measurement are fundamental. Choosing indicators, their weighting, defining norms, identifying objectives and stakes are specific problems for territories that are both complementary and competitive. More reliable spatio-temporal databases are needed for measuring the evolutions and comparing territorial dynamics.

A major issue is to adapt sources of information which were established for administrative or political units, at a given period in time, for their future use in evaluating territorial entities (cities, regions, networks) which have their own dynamics. The problem is crucial for long-term studies of the resilience and vulnerability of urban systems, or for a comparative evaluation of agenda 21 programs (which combine societal, economic and ecological objectives).

# 2.9. Ubiquitous computing

**Contributors:** Marc Schoenauer (INRIA)

**Keywords** : Peer to peer networks (P2P) , ad hoc networks, observation of multiscale spatio-temporal phenomena (trophic networks, agriculture, meteorology, ...), epidemic algorithms computational models and information theory , spatial computing, self-aware systems , common sens, privacy.

## Introduction

Today's technology makes it possible and even necessary to radically change the way we gather and process information, from the monolithic approach to the networked collaboration of a huge number of possibly heterogeneous comping units. This new approach should be based on distributed processing and storage, and will allow us to add intelligence to the many different artefacts which are increasingly important in our lives, and to compensate the foreseeable limits of classical Computer Science (end of the Moore era).

This long term objective requires :
- solving issues related to physical layout and communications (distributed routing and control)
- setting up self-regulating and self-managing processes
- designing new computing models
- specifying adaptive programming environments (using Machine Learning, retro-action and common sense).

It seems clear that we have today reached the technological limits of Von Neumann's sequential computational model. Hence new paradigms are necessary to meet the ever-growing demand for computational power of our modern society. The heart of these new paradigms is the distribution of computing tasks on decentralized architectures (e.g. multi-core processors and computer grids). The complexity of such systems is the price which must be paid to address the scaling and robustness issues of decentralized computing. Furthermore, it is now technologically possible to flood the environment with sensors and computing units wherever they are needed. However, an efficient use of widely distributed units can only be achieved through networking — and physical constraints limit the communication range of each unit to a few of its neighbours (ad hoc networks). At another scale, the concept of P2P networks also implies a limited visibility of the whole network. In both cases (ad hoc and P2P networks), the issue is to make an optimal use of the complete data which is available for the whole network. The challenges in this framework are targeted toward new computational systems, but will also address some issues raised in social or environmental networks, also treated in other pages of this road-map.

> **Main Challenges**
> 1. Local design for global properties (routing, control, confidentiality)
> 2. Autonomic Computing (robustness, redundancy, fault tolerance)
> 3. New computational models: distributing processing and storage, fusion of spatial, temporal and/or multi-modal data, abstraction emergence
> 4. New programming paradigms: creation and grounding of symbols (including proof and validation)

### 2.9.1. Local design for global properties

*Routing, control and privacy*

In order to better design and maintain large networks, we need to understand how global behaviours can emerge even though each element only has a very limited vision of the whole system, and makes decisions based on local information. A base model is that of epidemic algorithms, in which each element exchanges information with its neighbours only. The important issues are the type of information being exchanged (which should take into account privacy constraints) and the selection of corresponding neighbours. Both choices influence the global behaviour of the system.

**Methods**: Information theory; dynamical systems; statistical physics; epidemic algorithms; bio-inspired algorithms

### 2.9.2. Autonomic Computing

*Robustness, redundancy, fault tolerance*

The large-scale deployment of computational systems will not be possible without making those systems autonomous and, thereby, endowing them with properties of living systems such as natural robustness, reliability, resilience and homeostasis. However, the size and heterogeneity of such systems makes it difficult to come up with analytical models; moreover, the global behaviour of the system also depends on the dynamical and adaptive behaviour of the whole set of users.

**Methods**: Bio-inspired systems, self-aware systems.

### 2.9.3. New computing paradigms

*Distributed processing and storage, fusion of spatial, temporal and/or multi-modal data, abstraction emergence*

The networking of a large number of possibly heterogeneous computational units (grids, P2P, n-core processors) requires a huge computational power. However, in order to efficiently use such power, we need new computing paradigms that take into account the distribution of information processing on weak or slow units, and the low reliability of those units and of the communication channels. Similarly, data distribution (sensor networks, RFID, P2P) raises specific challenges such as integration, fusion, spatio-temporal reconstruction or validation.

**Methods**: Neuro-mimetic algorithms, belief propagation.

## 2.9.4. Specification of adaptive programming environments

*Machine learning, retro-action and common sense*

Programming ambient intelligence systems (domotic, aging, fitness) must include the user in the loop. The specification of the expected user behaviour requires a transparent link between the low-level data that are available and the user's natural concepts (e.g. symbol grounding). On the other hand, the research agenda must start by studying actual habits; such co-evolution of the user and the system leads to hybrid complex systems.

**Methods**: Brain Computer Interface, programming by demonstration, statistical learning.

# 2.10. Geosciences and the environment

**Reporter:** Michael Ghil (ENS Paris)

**Contributors:** Pierre Baudot (Inaf CNRS), François Daviaud (CEA), Bérengère Dubrulle (CEA), Patrick Flandrin (CNRS ENS Lyon), Cedric Gaucherel (INRA), Gabriel Lang (Agro Paris Tech), Francesco d'Ovidio (ENS), Daniel Schertzer (Météo France), Eric Simonet (CNRS).

**Keywords:** Climate change, predictability and uncertainties, ecosystems and landscapes, multiple scales and heterogeneity, climate and trophic networks, emergent diseases, transport and mixing, climate -weather interactions, stochastic vs. deterministic modeling.

## Introduction

The physical, chemical and biological environment of humans – from the local, community level to the global, planetary one – represents a rapidly increasing concern of the post-industrial era. Its study involves all the subsystems of the Earth system – the atmosphere, oceans, hydro- and cryosphere, as well as the solid Earth's upper crust – along with their interactions with the biosphere and with human activities. We are therefore dealing with a highly complex, heterogeneous and multiscale system, and with an exceedingly interdisciplinary set of approaches to it. The concepts and tools of complex-system theory seem particularly useful in attacking three major challenges. Firstly, the range of uncertainties still prevailing in future climate change projections has until now been attributed largely to difficulties in parameterising subgrid-scale processes in general circulation models (GCMs) and in tuning semi-empirical parameters. Recent studies also point to fundamental difficulties associated with the structural instability of climate models and suggest applying the theory of random dynamical systems to help reduce the uncertainties. Secondly, the Earth system varies at all space and time scales and is thus out of and probably far from thermodynamic equilibrium. The methods of statistical physics are therefore of interest in modeling the system's near-equilibrium behaviour and then extending the results farther away from equilibrium. Finally, much of the interest in this area arises from concern about the socio-economic impact of extreme events. The study of their statistics and dynamics can lead to a deeper understanding and more reliable prediction of these events.

The physical, chemical and biological environment of humans – from the local, community level to the global, planetary one – represents a rapidly increasing concern of the post-industrial era. The system's complexity is certainly comparable to that of systems studied in the life or cognitive sciences. It therefore appears highly appropriate to include this major area of applications of complex-system theory into the concerns of this road map.

The Earth system involves several subsystems – the atmosphere, oceans, hydro- and cryosphere, as well as the solid Earth's upper crust – each of which in turn is highly heterogeneous and variable on all space and time scales. Moreover, this variability is affected by and in turn affects the ecosystems hosted by each subsystem, as well as humans, their economy, society and politics. We are thus dealing with a highly complex, heterogeneous and multiscale system, and so the scientific disciplines needed to better understand, monitor, predict and manage this system are diverse and numerous. They include various subsets of the physical and life sciences, mathematics and informatics, and of course the full set of the geo-and environmental sciences, from geology, geophysics and geochemistry to the atmospheric and oceanic sciences, hydrology, glaciology and soil science.

Among the key interdisciplinary issues that arise in this major area are future climate change, change in the distribution of and interaction between species given past, present and future climate change, the way that the biogeochemical cycles of trace chemicals and nutrients interact with other changes in the system, and the connection between health issues and environmental change. On the methodological side, major objectives that would help to solve these issues include better prediction and reduction of uncertainties, better description and modeling of the transport and mixing of planetary fluids, understanding the net effect of weather on climate and the changes in weather as climate changes. Understanding the best uses of stochastic, deterministic or combined modeling in this highly complex setting is also essential.

To deal at the same time with some of these key issues and attempt to achieve some of the associated major objectives, we propose to focus on the following three main challenges: (i) to understand the reasons for and reduce the uncertainties in future climate change projections; (ii) to study the out-of-equilibrium statistical physics of the Earth system, across all scales; and (iii) to investigate the statistics and dynamics of extreme events.

The range of uncertainties in future climate change projections was originally determined in 1979 as an equilibrium response in global temperatures of 1.5–4.5 K for a doubling of atmospheric $CO_2$ concentration. After four IPCC assessment reports, it is still of a few degrees of end-of-century temperatures for any given greenhouse gas scenario. This persistent difficulty in reducing uncertainties has, until recently, been attributed largely to difficulties in parameterising subgrid-scale processes in general circulation models (GCMs) and in tuning their semi-empirical parameters. But recent studies also point to fundamental difficulties associated with the structural instability of climate models and suggest applying the theory of random dynamical systems to help reduce the uncertainties.

The Earth system varies at all space and time scales, from the microphysics of clouds to the general circulation of the atmosphere and oceans, from micro-organisms to planetary ecosystems, and from the decadal fluctuations of the magnetic field to continental drift. The entire system, as well as each of its subsystems, is a forced and dissipative system and is thus out of thermodynamic equilibrium and probably far away from it. The methods of statistical physics therefore seem of interest in modeling the system's near-equilibrium behaviour and trying to derive results that might then be extended to more realistic settings, farther away from equilibrium.

Finally, much of the interest in the geosciences and the environment arises from concern about the socio-economic impact of extreme events. The standard approach to such events rests on generalized extreme value theory (GEV). Its assumptions, however, are rarely met in practice. It is therefore necessary to pursue more sophisticated statistical models and to try to ground them in a better understanding of the dynamics that gives rise to extreme events. Based on better statistical and dynamical models, we should be able to provide more reliable predictive schemes for extreme events, and subject them to extensive testing across disciplines and data sets.

The geosciences have a long tradition of contributing to the study of nonlinear and complex systems. The work of E.N. Lorenz in the early 1960s has provided a major paradigm of sensitive dependence on initial state. His work and that of C. E. Leith have yielded deep insights into error propagation across scales of motion. Multiscale phenomena in the solid-earth and fluid-envelope context have helped refine the understanding of multi-fractality and its consequences for prediction across disciplines, even in the social and political sphere. We hope and trust that the work proposed here will prove equally inspiring and fruitful for the theory of complex systems and its applications in many other disciplines.

> **Main Challenges**
> 1. Understanding and reducing uncertainties.
> 2. Out-of-equilibrium statistical physics of the Earth system

## 2.10.1. Understanding and reducing uncertainties

Charney et al. (Natural Academic Press, 1979) were the first to attempt a consensus estimate of the equilibrium sensitivity of climate change in atmospheric CO2 concentrations. The result was the now famous range of 1.5K to 4.5K of an increase in global near-surface air temperatures Ts given a doubling of CO2 concentrations. Earth's climate, however, never was and probably never will be in equilibrium. In addition to estimates of equilibrium sensitivity, the four successive reports of the Intergovernmental Panel on Climate Change (IPCC: 1991, 1996, 2001, 2007) therefore focused on estimates of climate change over the 21st century, based on several scenarios of CO2 increase over this time interval. The general circulation models' (GCM) results of temperature increase over the coming 100 years have stubbornly resisted any narrowing of the range of estimates, with results for end-of-century temperatures still ranging over several degrees Celsius, for a fixed CO2 increase scenario. This difficulty in narrowing the range of estimates is clearly connected to the complexity of the climate system, the nonlinearity of the processes involved and the obstacles to a faithful representation of these processes and feedbacks in GCMs.

One obvious source of errors is the difficulty of representing all the processes that fall below the spatial and temporal resolution of the model. This problem is especially evident for biochemical processes, where the microphysical and microbiogical dynamics is coupled to the turbulent dynamics of the ocean and atmosphere and produces a spatiotemporal variability at virtually any scale of observation. One example is phytoplankton, whose fundamental role in absorbing CO2 is affected as much by the nutrient advection due to the large-scale circulation (basin scale, years), as by the presence of upwelling filaments (1-20 km, days), the ecological interaction with zooplankton (mm/m, hours/days), or the turbulent and biological processes at the cell scale. The study of such biochemical phenomena requires the development of novel theoretical tools that are beyond the capability of individual disciplines but which, because of their characteristics, fall naturally into the framework of complex systems. Such studies should be able to:
- deal at the same time with the various spatial and temporal scales of transport and tracer dynamics;
- integrate descriptions of different disciplines, notably transport and mixing properties from turbulence theory, and the biological and/or chemical processes of the advected tracer;
- provide results in a form that can be compared with ever-expanding observational datasets;
- and finally, allow to formulate a computationally-efficient parameterization scheme for circulation models.

A second source of errors lies in the fundamental difficulty related to the structural instability of climate models. It is well-known that the space of all deterministic, differentiable dynamical systems (DDS) has a highly intricate structure: the structural stable systems are unfortunately not typical of all deterministic dynamics, as originally hoped (Smale, 1967). Indeed, what is modeled by DDS does not appear to be typically robust from a qualitative, topological point of view, even for small systems like the Lorenz (1963) model. This disappointing fact has led mathematicians to grasp the problem of robustness and genericity with the help of new stochastic approaches (Palis, 2005). On the other hand, work on developing and using GCMs over several decades has amply demonstrated that any addition or change in a model's "parametrisations" - i.e. in the representation of subgrid-scale processes in terms of the model's explicit, large-scale variables - may result in noticeable changes in the model solution's behaviour.

The range of uncertainties issue, far from being a mere practical difficulty in "tuning" several model parameters, could be related to the inherent structural instability of climate models. A possible way of reducing this structural instability is the use of stochastic parametrizations with the aim of smoothing the resulting dynamics through ensemble average. A key question is then to determine whether ad-hoc stochastic parametrisations add some form of robustness to known deterministic climate models, and how they can reduce the range of uncertainties in future climate projections. Preliminary results indicate that noise has stabilizing effects that need to be investigated across a hierarchy of climate models from the simple to the most complex GCMs. Such an idea could be tested using theoretical concepts and numerical tools from the theory of random dynamical systems (RDS; L. Arnold, 1998). In this purely geometrical theory, noise is parametrised so as to treat stochastic processes as genuine flows living in an extended phase space called a "probability bundle". Random invariant sets such as random attractors can then be defined and rigorously compared, using the RDS concept of stochastic equivalence, thereby enabling us to consider the structural stochastic stability of these models.

## 2.10.2. Out-of-equilibrium statistical physics of the Earth system

The Earth and its various components (hydrosphere, atmosphere, biosphere, lithosphere) are typical out-of-equilibrium systems: due to the intrinsic dissipative nature of their processes, they are bound, without forcing, to decay to rest. However, in the presence of permanent forcing, a steady state regime can be established, in which forcing and dissipation equilibrate on average, allowing the maintenance of non-trivial steady states, with large fluctuations covering a wide range of scales. The number of degrees of freedom involved in the corresponding dynamics is so large that a statistical mechanics approach - allowing the emergence of global relevant quantities to describe the systems - would be welcome. Such a simplification would be especially welcome in the modeling of the fluid envelopes, where the capacity of present computers prohibits the full-scale numerical simulation of the (Navier-Stokes) equations describing them. Similar problems are ubiquitous in biology and environment, when the equations are known.

Another interesting outcome of a statistical approach would be to derive an equivalent of the Fluctuation-Dissipation Theorem (FDT), to offer a direct relation between the fluctuations and the response of the system to infinitesimal external forcing. Applied to the Earth system, such an approach could provide new estimates of the impact of climate perturbation through greenhouse gas emissions.

Various difficulties are associated with the definition of out-of-equilibrium statistical mechanics in the earth system, including:
- the problem of the definition of an entropy (possibly an infinite hierarchy of them) in heterogeneous systems;
- the identification of the constraints;
- the problem of the non-extensivity of the statistical variables, due to correlations between the different components of the system (possibly solved by introducing effective (fractional) dimensions).

On the physical side, several advances have been made recently in the description of turbulence, using tools borrowed from statistical mechanics for flows with symmetries. Variational principles of entropy production are also worth considering. Other advances have been made with regard to the equivalent of the FDT for physical systems far from equilibrium. Experimental tests in a glassy magnetic system have evidenced violation of the FDT through non-linearities in the relation between fluctuation and response. General identities between fluctuation and dissipation have been theoretically derived only for time-symmetric systems. They have been experimentally tested successfully in dissipative (non time-symmetric) systems like electrical circuits or turbulent flow. It would be interesting to extend these results to the Earth system.

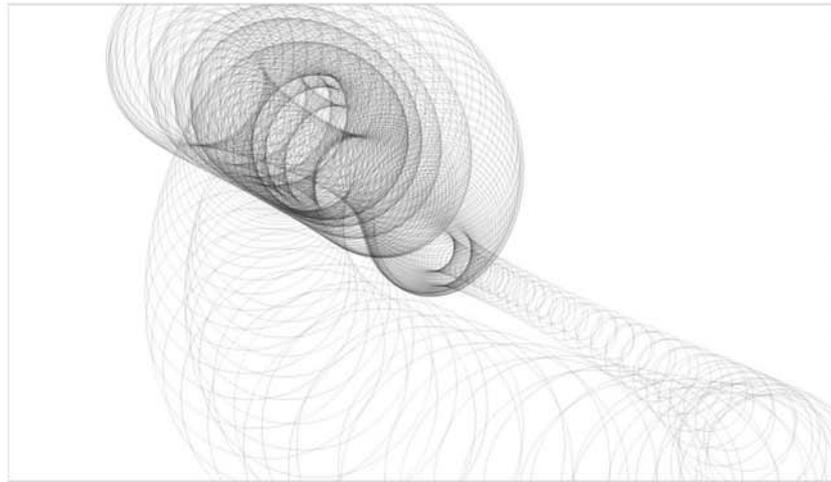

This second issue of the French Complex Systems Roadmap is the outcome of the "Entretiens de Cargèse 2008", an interdisciplinary brainstorming session organized over one week in 2008, jointly by RNSC, ISC-PIF and IXXI. It capitalizes on the first roadmap and gathers contributions of more than 70 scientists from major French institutions.

※ îledeFrance

**Editorial Committee**
Paul Bourgine – École Polytechnique, David Chavalarias – Institut des Systèmes Complexes de Paris Ile-de-France, Edith Perrier – Institut de Recherche pour le Développement

A "complex system" is in general any system comprised of a great number of heterogeneous entities, among which local interactions create multiple levels of collective structure and organization. Examples include natural systems ranging from bio-molecules and living cells to human social systems and the ecosphere, as well as sophisticated artificial systems such as the Internet, power grid or any large-scale distributed software system. A unique feature of complex systems, generally overlooked by traditional science, is the emergence of non-trivial superstructures which often dominate the system's behaviour and cannot be easily traced back to the properties of the constituent entities. Not only do higher emergent features of complex systems arise from lower-level interactions, but the global patterns that they create affect in turn these lower levels — a feedback loop sometimes called immergence. In many cases, complex systems possess striking properties of robustness against various large-scale and potentially disruptive perturbations. They have an inherent capacity to adapt and maintain their stability.

Because complex systems require analysis at many different spatial and temporal scales, scientists face radically new challenges when trying to observe complex systems, in learning how to describe them effectively, and in developing original theories of their behaviour and control.

The aim of this roadmap is to identify a set of wide thematic domains for complex systems research over the next five years. Each domain is organized around a specific question or topic and proposes a relevant set of "grand challenges", i.e., clearly identifiable problems whose solution would stimulate significant progress in both theoretical methods and experimental strategies.

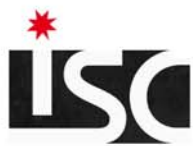
iscpif.fr

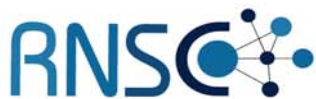
rnsc.fr

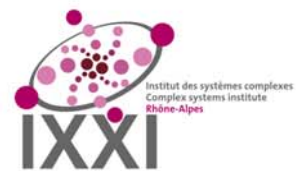
ixxi.fr